\documentclass[draft]{agujournal2019}
\usepackage{subcaption}
\usepackage{comment}
\usepackage{lineno}
\usepackage{soul}
\usepackage{amsmath}
\usepackage{url}
\newcounter{SuppCount}
\newcommand{\SuppRef}[1]{%
	\stepcounter{SuppCount}%
	#1\theSuppCount
}
\newcommand{\iset}{\mathcal{D}_L}
\newcommand{\bandiset}{\Delta^{L_1,L_2}}
\newcommand{\Q}{Q_{\rm tot}}
\newcommand{\dd}{\mathrm{d}}
\newcommand{\rhoref}{\bar\rho_{\rm ref}}

\draftfalse

\journalname{QJRMS}

\begin{document}

\title{Upscale energy transfer from mesoscale convective systems to African easterly waves}
\authors{Ben Maybee\affil{1}, Elliot McKinnon--Gray\affil{2}, Dan Shipley\affil{2}, and John Methven\affil{2}}

\affiliation{1}{Institute for Climate and Atmospheric Science, University of Leeds, Leeds, UK}
\affiliation{2}{Department of Meteorology, University of Reading, Reading, UK}
\correspondingauthor{Ben Maybee}{b.w.maybee@leeds.ac.uk}


\vspace{2cm}
\begin{abstract}
    African easterly waves (AEWs) drive the synoptic variability of West African monsoon rainfall via intricate two--way interactions with mesoscale convective systems (MCSs), large organised complexes of moist deep convection. The composite MCS--AEW system is a canonical example of multiscale interactions in the tropical atmosphere, with upscale feedbacks from deep convection driving the evolution of a larger--scale circulation. Multiple authors show MCS feedbacks play an integral role in supporting AEW growth and maintenance. Here we support this perspective through application of techniques from turbulence theory, calculating for the first time patterns of interscale kinetic energy transfer, $\iset$, across different spatial scales $L$ around MCSs and AEW lifecycle phases. We study AEW and MCS dynamics in two pan--tropical 40 day Met Office Unified Model simulations with $\sim$5\,km grid spacings, for which $\iset$ is calculated with the LoSSETT software package. We find explicitly--simulated MCSs show a distinctive vertical dipole structure of $\iset$ at scales below 1000\,km, with strong upscale energy transfer located in the upper troposphere within storm anvils. Regionally, upscale energy transfer occurs most strongly where organised convection is active. From composite and Hovm\"{o}ller analysis we find 200\,hPa upscale transfer consistently leads AEW troughs, and that growing continental AEWs in the explicit simulation are characterised by strengthened upscale transfer of energy from MCS to AEW scales. Case--study analysis shows this transfer can contribute to the eddy kinetic energy growth of strengthening AEW troughs. These signals are weak in a simulation with parameterised convection: here, zonal mean energetics show enhanced barotropic conversion, caused by a stronger African Easterly Jet, provides a stronger AEW population. Our analysis of upscale energy transfers complements, and is consistent with, previous analysis of AEW potential vorticity dynamics, providing a route towards building new theoretical understanding of tropical dynamics using km--scale models.
    \vspace{2cm}
\end{abstract}


\section{Introduction}
\label{s:Intro}

African Easterly Waves (AEWs) play a key role in conceptual understanding of tropical atmospheric dynamics \cite{Tomassini2020interaction}: they are a canonical example of moist deep convection playing a key active, ``upscale'' impact on a large--scale circulation, with convective heating fundamental to AEW growth and maintenance \cite{Hsieh2005generation,Tomassini2017interaction,Russell2020potential,Adames2026rotation}. See also extensive reviews by \citeA{Parker2017meteorology} and \citeA{Nunez2026review}. As the primary mode of boreal--summer synoptic variability over sub--Saharan, northern--hemisphere Africa \cite{Reed1977structure}, AEWs are a key component of the West African monsoon circulation. They exist as barotropic--baroclinic disturbances of the African Easterly Jet (AEJ), which has tight--enough curvature for a reversal of the meridional potential vorticity gradient \cite{Pytharoulis1999low,Kiladis2006three}. Waves are found to propagate at different levels on either flank of the AEJ, forming 650\,hPa southern and 850\,hPa northern trains with distinct characteristics \cite{Thorncroft2001african,Duvel2021vortices}. Southern--track AEWs have roughly 3--5 day periods, 2000--5000\,km wavelengths and 8--12\,ms$^{-1}$ phase speeds \cite{Nunez2026review}. Northern--track waves have longer periods (6--9 days), shorter wavelengths, and slower phase speeds \cite{Diedhiou1999easterly,Jonville2025distinguishing}. It is the southern--track AEWs which exhibit strong couplings to mesoscale convective systems \cite{Semunegus2017characterization}, with these waves playing a significant downstream role as the seed vortices for Atlantic tropical cyclogenesis \cite{Landsea1993climatology,Russell2017revisiting}.

Mesoscale convective systems (MCSs) are large, organised clusters of thunderstorms that, in the tropics, cause around 45\% of mean rainfall \cite{Feng2025mesoscale} and 80\% of extreme 100\,mm/day rainfall accumulations \cite{Roca2020extreme}. For detailed reviews, see \cite{Houze2004mesoscale} and \citeA{Schumacher2020formation}. In atmospheric dynamics, their key attribute is an internal mesoscale circulation which couples together convective and stratiform heating regions, leading to multi--modal vertical heating and mass flux profiles. Strong convective updrafts drive strong heating and thus diabatic PV generation in the lower troposphere \cite{Tomassini2017interaction,Russell2020potential}. In the squall line--type MCSs typically found over West Africa, these features are then followed by significant regions of upper--level heating that generate positive mid--level and negative upper--level PV anomalies \cite{Johnson2020potential}. This stratiform component can interact with and influence circulations which extend far beyond the MCS scale. For example, upper--level generation of negative PV by North American MCSs under certain large--scale flow patterns leads to the modification of Rossby wave packets and eventual downstream Rossby wave breaking \cite{Parsons2019role,Lojko2022remote}, which can form blocking highs over Europe \cite{Rodwell2013characteristics}. 

AEWs and MCSs are found to dance complex, two--way feedback loops. From a ``downscale'', AEW--to--MCS perspective, AEWs influence storms' large--scale convective environment \cite{Janiga2014convection}, with favourable strong shear, high moisture regions associated with the leading edge of wave troughs and providing predictability for convective rainfall \cite{Bain2014objective}. Our primary interest here is the ``upscale'' MCS--to--AEW perspective. Many dynamical mechanisms contribute to (southern--track) AEW growth, some features of which can be described from dry or moist standard baroclinic and barotropic instability analysis \cite{Thorncroft1994idealized,Thorncroft1999maintenance,Cornforth2009impact}. A complete study of the regional energetics of AEWs by \citeA{Hsieh2007study} showed that barotropic and baroclinic energy conversion both provide significant contributions to the generation of AEW eddy kinetic energy (EKE). Their Lorenz energy--cycle techniques have provided the platform for numerous insightful studies of AEW behaviour, providing the process analysis techniques for explaining the response of AEWs to increased atmospheric moisture \cite{Nunez2024moisture}, orographic forcings \cite{Hamilton2020topographic}, and climate change \cite{Akinsanola2025projected}.

Modern perspectives centred in AEW potential vorticity budgets additionally emphasise the importance of MCS convection, due to its role in diabatic PV generation. \citeA{Berry2005case} and \citeA{Thorncroft2008three} argued that in this manner, MCSs could contribute to AEW initiation, maintenance and propagation, while \citeA{Hsieh2005generation} proposed convective activity enhanced large--scale baroclinicity which thereby faciliated wave growth. Using a long term study of ERA5 reanalysis, \citeA{Russell2020potential} showed that diabatic PV generation is the dominant term in AEW growth, with eddy advection by the background flow then playing the key role in wave propagation. Sensitivity experiments provide further causal support for this feedback. \citeA{Berry2012african} and \citeA{Russell2020african} have independently shown that suppressing convective heating in propagating AEWs leads to rapid decay of the wave trough that does not occur if convection is maintained. \citeA{Tomassini2017interaction} used PV tracers in high--resolution experiments to identify how diabatic PV generation sustains AEW circulations, and to show that misrepresentation of convective heating leads to misrepresented wave dynamics. Separately, AEWs which couple to MCSs are more likely to undergo downstream cyclogenesis, through both onshore diabatic PV reinforcement and offshore vortex stretching \cite{Schwendike2010convection,Leppert2013relation,Nunez2020wave,Russell2020african}.

The horizontal scale of MCSs is much smaller than AEW wavelengths: typical ranges put MCS anvils near the bottom of the meso--$\alpha$ spectrum (around $\mathcal{O}$(400\,km)), and the cyclonic vortices that comprise AEW troughs towards the top (around $\mathcal{O}$(1500\,km)). How do the smaller scale convective eddies come to drive the large--scale synoptic eddy? The upscale nature of this interaction was particularly emphasised by \citeA{Tomassini2017interaction} and \citeA{Russell2020african}. We have been loosely using the word ``upscale'' to describe the influence of convection on a much larger scale circulation, but it carries a formal definition in turbulence theory, where it refers to the interscale transfer of energy across a scale $L$ to larger scales. In two--dimensional vortex dynamics, energy is transferred upscale at every wavenumber $k$ in the global turbulent cascade above a forcing scale, whereas in three--dimensions, vortex--shear and shear--shear interactions cause global downscale energy transfer \cite{Salmon1998lectures,Johnson2020energy}. However, within this global, statistical statement, upscale energy transfer over large regions of the flow can occur (e.g.~\citeA{Vieweg2022inverse}). This principle has been used in oceanography to explain the crucial, upscale role that mesoscale eddies play in driving ocean currents and overturning circulations \cite{aluie2018JPO,storer2023}. Several authors have established the foundations for conducting similar, local interscale energy transfer analysis in the atmosphere \cite{Faranda2018computation,kouhen2024,Sichani2026atmospheric}. Tantalisingly, \citeA{Mckinnon2025representation} show upscale energy transfer is stronger in convection--permitting simulations of equatorial waves that explicitly simulate deep convective motion, as opposed to relying on a convection parameterisation scheme. Strengthened upscale energy transfers may contribute towards the more realistic simulation of convectively--coupled waves in convection--permitting models \cite{Judt2021resolved,Mckinnon2025representation}.

The purpose of this paper is to apply the formal understanding of upscale energy transfer available from turbulence theory to the multiscale interactions between MCSs and AEWs. By calculating the local interscale transfer of kinetic energy in high resolution simulations over West Africa, we offer a novel perspective on the dynamical feedbacks between convective and synoptic--scale eddies already documented in the literature. Our results provide understanding of the key role MCSs play in AEW growth, demonstrating that MCSs can act as a source of upscale energy transfer to AEW scales.  We begin by describing the two equal--resolution km--scale simulations on which our results are based, and providing details of the calculation of interscale energy transfer $\iset$ from the LoSSETT software package \cite{LoSSETT}. In Section~3 we show that MCSs cause upscale energy transfer in the upper troposphere and explore their influence on the regional energy spectrum, before considering the role of MCS $\iset$ in AEW lifecycles in the explicit convection simulation (Sec.~4). In Section~5 we show that upscale energy transfer is not essential to strong AEW development, with barotropic and baroclinic energy conversion explaining stronger AEW amplitudes found in the simulation with parameterised convection. However, this is only because of significant biases in West African climate stemming from the parameterisation of convection. We therefore conclude in Sec.~6 that upscale energy transfer from MCSs contributes to AEW growth, and likely explains the importance of MCS diabatic heating to the evolution of observed waves.


\section{Data and Methods}

\subsection{Simulations}

Our study is based on analysis of two atmosphere--land runs of the km--scale grid  Cyclic Tropical Channel configuration of the Met Office Unified Model (MetUM, \citeA{brown2012UM}). Both simulations share a common domain spanning all longitudes and -40$^\circ$---26$^\circ$N of latitude, an N2560 grid spacing equivalent to 8$\times$5\,km at the equator, and use 90 vertical levels up to 40\,km. The simulations are both initialised from MetUM operational analysis and span 40 days starting from 2016--08--01, thus aligning with the summer period of the DYAMOND multi--model intercomparison project \cite{stevens2019dyamond}. Latitudinal boundary conditions are derived from an N1280 global simulation initialised from the same analysis; lower boundary sea surface temperature is updated daily using 1/20$^\circ$ Operational Sea Surface Temperature and Sea Ice Analyis (OSTIA) data \cite{donlon2012OSTIA}. Both simulations output hourly single--level diagnostics and 3--hourly multi--level fields.

The key difference between the simulations is their treatment of convection. ``Explicit'' utilises the RAL3.2 convection--permitting science configuration \cite{bush2025RAL3}, whereas ``Param'' uses the GAL9 configuration which features a CAPE--closure mass flux parameterisation scheme \cite{willett2026met}. The simulations already underpin several studies of tropical dynamics \cite{jones2023}. Explicit has been shown to simulate more realistically West African MCS physics and environmental responses (``Trop5--exp'', \citeA{Maybee2024wind}); the distribution of total column water vapour over tropical oceans \cite{Bassford2026tropical}; and characteristics of equatorial wave modes (``RAL'', \citeA{Mckinnon2025representation}). In this study, we focus on a domain spanning 0$^\circ$---25$^\circ$N, -50$^\circ$---30$^\circ$E, thereby covering most of the African continent between the Sahara desert and Gulf of Guinea, and a large extent of the Atlantic ocean to the west. Analysis of mesoscale vortex dynamics (including AEWs) over a similar domain in global, year--long MetUM simulations using the same science configurations is available in \citeA{Maybee2026dynamics}.

Based on this previous work, we can already state several key differences between the simulations which provide important context to our results. Meridional gradients of boundary layer temperature and moisture are stronger in Param than Explicit, with higher maximum zonal mean values, and the temperature gradient driving a significantly stronger mid--level AEJ that lies several degrees farther south (Fig.~S\SuppRef{}). This in part stems from a diurnal cycle of convection which peaks unrealistically early \cite{jones2023}, in common with most parameterised simulations globally, which in West Africa prevents appropriate ventilation of the Sahara by MCS cold pools \cite{Marsham2013role,Birch2014seamless}. The MCSs in Explicit show more realistic responses to their environment, particularly low--level vertical wind shear, and are more intense than in Param, with colder cloud tops and heavier rainfall rates \cite{Maybee2024wind}. Param therefore underestimates rainfall over the Sahel. However, in common with other CP models \cite{Feng2025mesoscale} the convective intensity of Explicit MCSs are overestimated versus observations, and near the coast and offshore, in particular, storm anvil sizes are significantly underestimated. In the global, longer--run cousins of our simulations the number of AEWs is equivalent between science configurations, but their evolution and tracks differ \cite{Maybee2026dynamics}.

Since Explicit offers a more realistic representation of the West African monsoon system, it will be the primary focus of our paper, which is interested in MCS--AEW interactions. Results from Param are intended to add context and provide further physical insight into dynamical processes, rather than sit as an equally--weighted model comparison.

\subsection{Upscale energy transfer}

To diagnose interscale energy transfers, we use the tools developed for and described in \citeA{Mckinnon2025representation}, in particular the Local Scale--to--Scale Energy Transfer Tool, LoSSETT \cite{LoSSETT}. Computation and analysis in \citeA{Mckinnon2025representation} used the same Explicit and Param simulations analysed in this paper. LoSSETT computes local interscale energy transfer as a function of space, time, and spatial scale $L$ using the descriptions of inertial energy dissipation in three--dimensional incompressible fluids derived by \citeA{Duchon2000inertial}. This theoretical framework forms the basis for a previous investigation of local atmospheric interscale energy transfer using global reanalysis \cite{Faranda2018computation}. For our analysis of AEWs, it suffices to note that \citeA{Faranda2018computation} show that the kinetic energy $K_L$ at length scales equal to and larger than $L$ satisfies
\begin{equation}
    \frac{\partial K_L}{\partial t} + \nabla\cdot\mathbf{J}_L = -\varepsilon_L+{Q}_L-\mathcal{D}_L\,.
    \label{eqn:iset_budget}
\end{equation}
Here $\mathbf{J}_L$ is the flux of KE on scales larger than $L$; $\varepsilon_L$ is dissipation at scale $L$; ${Q}_L$ is the conversion of available to kinetic energy; and $\iset$ is the interscale transfer of energy across a length scale $L$, with positive values indicating transfer downscale to scales smaller than $L$. All of these quantities are \textit{local} in both space and time. 

Calculation of $\iset$ is based on velocity increments $\delta\mathbf{u}$ \cite{Faranda2018computation} and requires specification of a filter kernel with characteristic length scale $L$ to pick out the scales of interest. To facilitate standard atmospheric analysis, LoSSETT calculates $\iset$ over a 2D kernel on a spherical surface of radius $R$, $S^2(R)$, assumed to be orthogonal to Earth's gravity (i.e. Earth is assumed spherical). In this special case,
\begin{equation}
    \iset= R\!\int_0^{\pi R}\!\!\dd r\, \frac{\partial G_L(r)}{\partial r}\sin\!\left(\frac{r}{R}\right) \left[\int_0^{2\pi}\!\!\dd\chi\,  \hat{\mathbf{r}}\cdot\delta\mathbf{u}\,|\delta\mathbf{u}|^2\right]\,,
    \label{eqn:iset}
\end{equation}
where the velocity increment $\delta\mathbf{u}\equiv\mathbf{u}(\mathbf{x}_2)-\mathbf{u}(\mathbf{x}_1)$, $\mathbf{x}_1$ and $\mathbf{x}_2$ are two points on $S^2(R)$ subtending an angle $\chi$ at the centre of $S^2(R)$, $r = R \chi$ is the great circle (geodesic) distance between $\mathbf{x}_1$ and $\mathbf{x}_2$, $\hat{\mathbf{r}}$ is the tangent vector to the great circle joining $\mathbf{x}_1$ and $\mathbf{x}_2$, and $G_L$ is a compactly supported ``standard mollifier'' filter kernel \cite{kuzzay2015},
\begin{equation}
    G_L \equiv 
    \begin{cases}
        \mathcal{N}_L\exp\left(-\frac1{1-(r/L)^2}\right) & r\leq L\\ 0 & r>L
    \end{cases}\,.
\end{equation}
For more general forms of $\iset$, see \citeA{Mckinnon2025representation} and \citeA{Faranda2018computation}. In practice, LoSSETT first calculates velocity increments and then calculates the angular average in square brackets in Eq.~\ref{eqn:iset} taken over the local spherical cap centred at $\mathbf{x}_1$ using a quadratic approximation to spherical geometry. In tropical regions there is negligible error versus full spherical geometry \cite{Mckinnon2025representation}.

Since $\iset$ is computed using a filter kernel which covers a larger spatial domain with increasing $L$, the computational cost of calculations scales strongly with increasing $L$. In practice, this limits the maximum $L$ that is computationally feasible when computing from native grid N2560 data. For upscale energy transfers that may contribute to AEW dynamics, we expect meso--$\alpha$ (200---2000\,km) scales to be most relevant: AEW wavelengths are of order 4000\,km, while it is organised convective systems which are found to influence AEW dynamics, not scattered convective updrafts. For our purposes it is therefore reasonable to coarsen the native grid data prior to application of LoSSETT, accepting that we cannot make statements about near grid--scale $\iset$ (i.e. meso--$\gamma$) energy transfers. The pay off is that we can efficiently compute $\iset$ out to AEW scales. We find that using N640 resolution ($\sim$25\,km at the equator) winds, for $L$ larger than 200\,km there is negligible difference to $\iset$ values calculated at native resolution (Fig.~S\SuppRef{}). The smallest computed scale in this output is $L=64$\,km, and the largest is $L=5000$\,km. In practice we will not consider scales below 100\,km. 

For grid points which are less than $L$\,km from a CTC latitudinal boundary, calculations are undertaken using an embedding of the CTC within its driving model (N1280 global GAL9 simulation) whereby driving model fields are used to fill the extension of the filter kernel beyond 26$^\circ$N. This ensures we can make robust statements about large--$L$ value energy computations for AEWs, which lie approximately 1500\,km south of the northern CTC boundary. In all plots of interscale transfer $\iset$ across a given scale $L$, negative (blue) values represent upscale transfer. We also choose to weight the interscale energy transfer by a reference density profile $\rhoref$, which is a function of pressure only, since stronger upper--atmospheric winds otherwise dominate vertical profiles. Such mass--weighting continues to respect all conservation laws. The reference profile is calculated as the mean $\bar\rho(p)$ over the full domain of interest for each simulation individually, and only acts to impose a common scale upon vertical profiles of $\iset$.

As discussed in the Introduction, we expect regions of upscale energy transfer to be collocated with regions of diabatic heating. To study this, here the apparent atmospheric heating diagnostic $Q_{\rm tot}$ is calculated from coarse 0.5$^\circ$ fields, where the heating at a given pressure level is calculated as \cite{Luo1984large}
\begin{equation}
    Q_{\rm tot} \equiv c_p \frac{DT}{Dt} = c_p\frac{\partial T}{\partial t} + c_p \mathbf{v}\cdot\nabla T + c_p \left(\frac{p}{p_s}\right)^\kappa\omega\frac{\partial\theta}{\partial p}\,,
\end{equation}
where $\theta$ is potential temperature, $\omega$ vertical pressure tendency, $p_s$ the surface pressure and $\mathbf{v}$ the horizontal velocity components. The exponent $\kappa = R_d/c_p$, where $R_d$ is the specific gas constant and $c_p$ the specific heat capacity at constant pressure, both for dry air. $Q_{\rm tot}$ represents the larger--scale response to heating from an apparent source (typically labelled $Q_1$; \citeA{Tao1993heating,Olson1999atmospheric}), which depends on sub--grid radiative and microphysical processes that we cannot quantify from available model diagnostics. We therefore only use coarsened model fields at much larger scales than the grid--scale to calculate $Q_{\rm tot}$.

\subsection{AEW and MCS tracking}
\label{ss:AEWMCStracking}

As noted in the Introduction, our focus is on southern--track, mid--level 3--5 day period AEWs, which couple strongly to MCS dynamics. To objectively identify AEW troughs we use the QTrack algorithm described in \citeA{Lawton2022influence}, and compared against other objective tracking methods in \citeA{Nunez2026review}. QTrack identifies and tracks maxima in the 700\,hPa curvature vorticity (CV) field computed from the non--divergent component of 1$^\circ$--coarsened 6 hourly wind fields. To identify CV maxima, the average CV within a 600\,km radius of each grid point is computed, and wave trough centres are identified using weighted centroids. Over land wave identification uses meridional averages of CV, following the method of \citeA{Elless2018african}, while over ocean forward--estimation of trough position from AEW velocity is used, following \citeA{Brammer2015variability}. In both cases the tracker identifies a central coordinate of the AEW trough and its strength, but does not identify a full trough axis (as in e.g.~\citeA{Fischer2023objective}). Here we use the QTrack trough locations and strengths as the primary measures of the AEW populations in our simulations, with tracking conducted over the full study domain.

In addition to identifying AEW signals through an objective tracker, we also adopt the common technique of spectral filtering to isolate signals with 3--5 day periodicity. We isolate this periodicity range by applying a wider 2--8 day Mexican Hat wavelet filter, deliberately using a wider filter to include signals which overlap into the window of interest; we also do not seek to entirely exclude northern--track wave signals, as is sometimes required. AEWs can then be identified as synoptic eddies in the filtered mid--level winds, $u' = u - \bar{u}$, where $\bar{u}$ is the simulation mean (either temporal or spatial; \citeA{Diaz2013energy,Hsieh2007study}) and all fields are filtered. The eddy kinetic energy (EKE) is then
\begin{equation}
    K_e(t) = \frac12{\left({u'}^2(t) + {v'}^2(t)\right)};\label{eqn:eke}
\end{equation}
note that this kinetic energy measure does not explicitly satisfy Eq.~\ref{eqn:iset_budget}.

For objective identification of MCSs we use the simple snapshot--detection method developed in \citeA{Klein2021combining} and previously used to study MCSs in our simulations \cite{Maybee2024wind}. MCSs are identified hourly over a continental subdomain (5$^\circ$--20$^\circ$N, -18$^\circ$--25$^\circ$E) from 16--21 UTC as contiguous snapshots of brightness temperature $T_b<$ -50$^\circ$C covering at least 5000\,km$^2$ and exhibiting maximum rainfall $>$1 mm hr$^{-1}$. Top--of--atmosphere brightness temperature $T_b$ is calculated from outgoing longwave radiation via the empirical conversion of \citeA{Yang2001diurnal}. The selected hours are the diurnal peak of MCS activity in our domain, and we restrict to this period to enable consistent study of mature MCS responses to their preceeding convective environment. Environmental conditions are calculated as the 12 UTC mean conditions across a 0.75$^\circ$ box centred on the location of minimum $T_b$, representing an undisturbed pre--convective environment \cite{Klein2021combining}. We find no change to our results when using a more sophisticated tracking algorithm that captures the full lifecycle of individual MCSs (described in \citeA{Maybee2025how}).


\section{Upscale energy transfer, MCSs, and length scales}
\label{sec:mcs}

\begin{figure}[t]
    \centering
    \includegraphics[width=\textwidth]{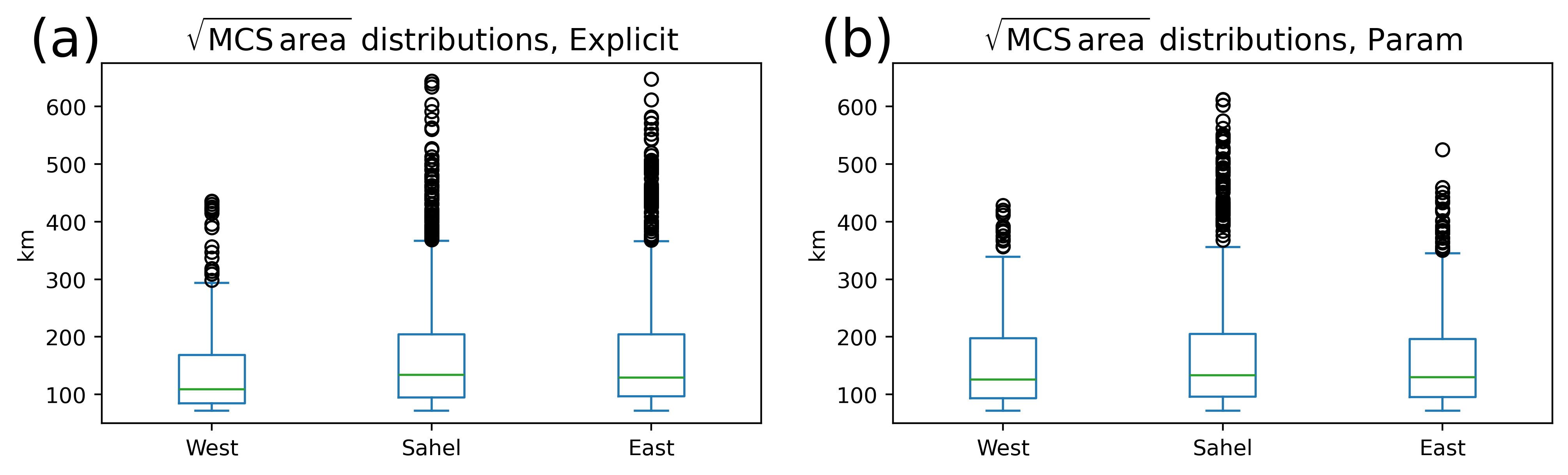}
    \includegraphics[width=\textwidth]{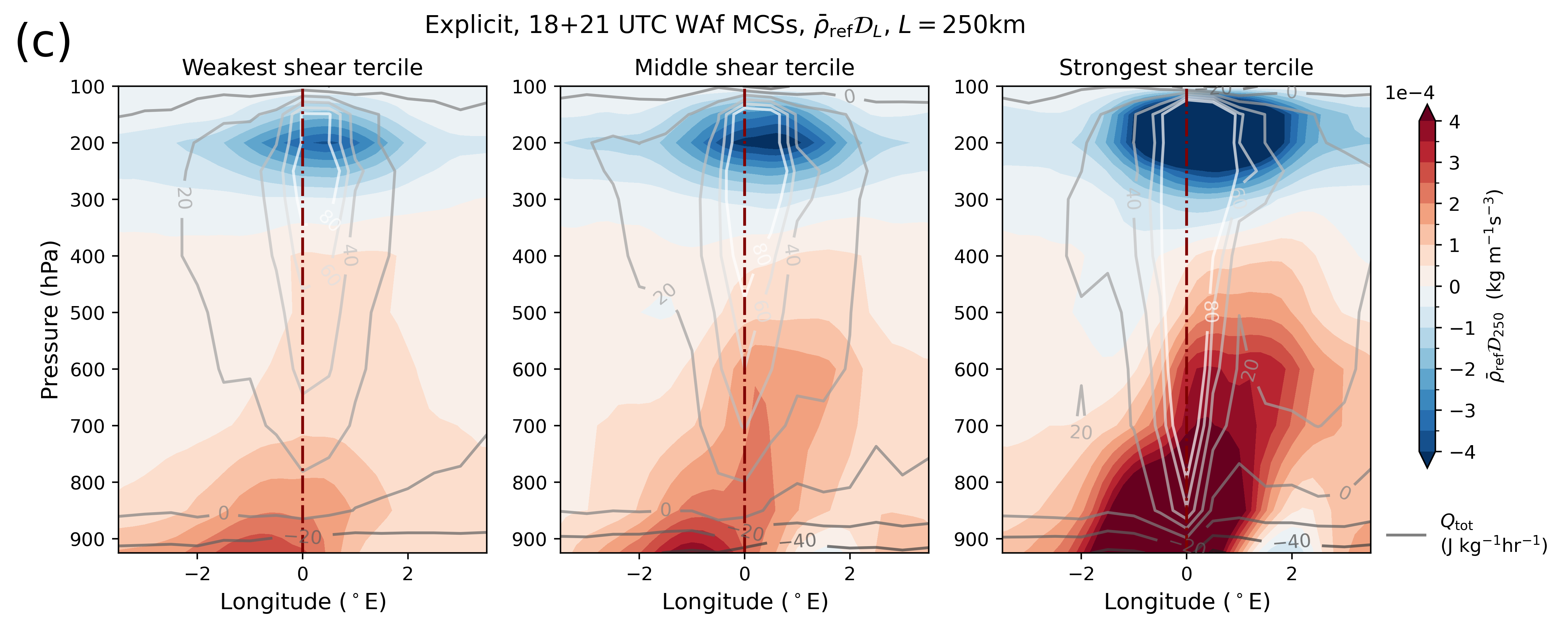}
    \caption{\textit{(a)} Distributions of the square--root of Explicit 16--21 UTC MCS anvil areas, separated into spatial bins based on the location of system minimum $T_b$. West MCSs are those between $18^\circ$--10$^\circ$W, Sahel those between -10$^\circ$--10$^\circ$E, and East those between 10$^\circ$--25$^\circ$E. Whiskers extend to 1.5$\times$ the interquartile range. \textit{(b)} Repeated for Param MCSs. \textit{(c)} Zonal vertical composites of $L$=250km mass--weighted interscale energy transfer $\iset$ (shading), and apparent atmospheric heating (contours), sampled about 18 and 21 UTC West African MCS locations of minimum $T_b$ in Explicit. The plotted MCS sample has limited time sampling due to three--hourly vertical wind diagnostics, and is split across panels by tercile of environmental 12 UTC 650--850hPa zonal wind shear. Latitudinal averages are taken over a 3$^\circ$ central window.}
    \label{fig:mcsISETcomps}
\end{figure}

MCSs are frequently attributed with upscale effects whereby they influence the evolution of larger--scale circulation patterns, with AEWs a longstanding canonical example. The LoSSETT calculation of interscale energy transfer provides a route towards formalising this statement at a length scale $L$. A key starting point is identifying the relevant scale, $L$. MCSs are themselves comprised of intricate internal scale--interactions, and perhaps the most important physics driving systems' mesoscale circulations is that of the km--scale convective updraft cores. Here however, we are interested in the transfer of energy from MCSs to larger scales, for which we expect stratiform heating to be most relevant \cite{Schumacher2004tropical,Russell2020african}. As an indicator of relevant length scales we therefore plot distributions of the square roots of MCS areas in Explicit (Fig.~\ref{fig:mcsISETcomps}a) and Param (Fig.~\ref{fig:mcsISETcomps}b). Using coordinates of MCS minimum $T_b$ to group storms into longitudinal bins, we find MCS areas increase moving east across Africa in Explicit, with offshore systems in particular considerably smaller than those in central latitudes. Area distributions in Param are skewed towards larger values, but for both simulations and all spatial bins the median $\sqrt{\rm MCS\,area}$ value is around 150\,km, and boxplot whiskers extend to near 400\,km. Since MCSs do not have isotropic geometries, we conclude that over West Africa the relevant MCS range of $L$ is $\sim$100---400\,km.

Figure~\ref{fig:mcsISETcomps}c then shows vertical profiles of composite mass--weighted $\iset$ for $L$=250\,km centred on mature West African MCSs in Explicit at the diurnal peak of convection (evening). In all panels, the pattern of $\iset$ shows a dipolar pattern, with downscale energy transfer (red shading) in the boundary layer and upscale energy transfer (blue) at the tropopause. The magnitude of upscale energy transfer is stronger than that of the downscale transfer, even with the density profile scaling. The dipoles reflect the vertical structure of the horizontal divergence field typically found around MCS convective cores, with convergence of low--level flow into a storm updraft eventually leading to divergent detraining outflows in the upper atmosphere. This dipole of upper level upscale, lower-level downscale transfer seems to be a robust feature of convective systems \cite{Green2020resolved,Mckinnon2025representation}. We find that strong upper--level outflows from MCS cores into their anvil shields and beyond yield upscale energy transfer for scales $L$ from 64 to 2000\,km. 

In Fig.~\ref{fig:mcsISETcomps}c, panels are split into terciles of environmental low--level vertical wind shear (shear henceforth) values encountered by regional MCSs, where the shear is taken as the 12 UTC difference between 650 and 850\,hPa zonal winds \cite{Maybee2024wind}. We find that upscale energy transfer increases with shear tercile: MCSs in the strongest shear conditions yield the strongest values of $\iset$ (of both signs). This finding does not depend on our choice of $L$, and also holds for environmental total column moisture (Fig.~S\SuppRef{}a). It is well established that dynamical environmental controls such as vertical shear, and thermodynamic controls such as moisture, enhance MCS convection, with strengthening of either control observed to cause colder, more intense systems with higher rainfall and larger anvils \cite{Janiga2014convection,Taylor2017frequency,Klein2021combining,Roca2025scale}. Interscale energy transfer therefore behaves in a similar manner to other measures of convective intensity, occurring as a dynamical consequence of convective activity that is increased by favourable environmental conditions. Thus in these environments, MCSs generate stronger upscale energy transfers and have a stronger upscale influence. Note that the (upshear) tilts evident in all panels are primarily due to the horizontal vorticity balance between the low--level wind shear and MCS cold pools \cite{Rotunno1988theory}, rather than a signature of convection embedded within a baroclinic wave. The median separation distance between a member of the plotted MCS sample and a tracked AEW trough is 1016\,km, while MCS composites from a previous cold pool suppression experiment show a reversal of the updraft tilt \cite{Maybee2025how}.

Contours of $\Q$ further show that the increase in $\iset$ under stronger environmental shear (and moisture) follows that of the convective heating: going from lowest to highest shear tercile, the heating profiles strengthen, deepen, and extend further into the stratiform region trailing (east of) the convective updrafts. Diabatic PV generation will also therefore increase under more favourable conditions and thus behaves consistently with $\iset$; we will return to correlations between PV and $\iset$ structures in Sec.~\ref{sec:mcs}. To the east of the composites centres in Fig.~\ref{fig:mcsISETcomps}c there is boundary layer cooling caused by cold pools from MCS downdrafts, which are found to have downscale energy transfers; this is a surprising result given that in the Sahel, cold pools can spread to enormous size \cite{Provod2016characterization,Hoeller2024characteristics} and influence regional climate \cite{Maybee2025how}. Figures.~S3b,c also show equivalent $\iset$ results for parameterised MCSs at $L$=500\,km: we find  mean values of $\iset$ approximately five times weaker than in Explicit, and only a limited influence of shear on $\iset$ and heating profiles, reflecting the unrealistic control of shear on storm characteristics in this simulation \cite{Maybee2024wind}. Moreover, we find no characteristic dipole signature at $L$=250\,km, despite simulated MCSs having comparable scales to Explicit.

\begin{figure}[h!]
    \centering
    \includegraphics[width=\textwidth]{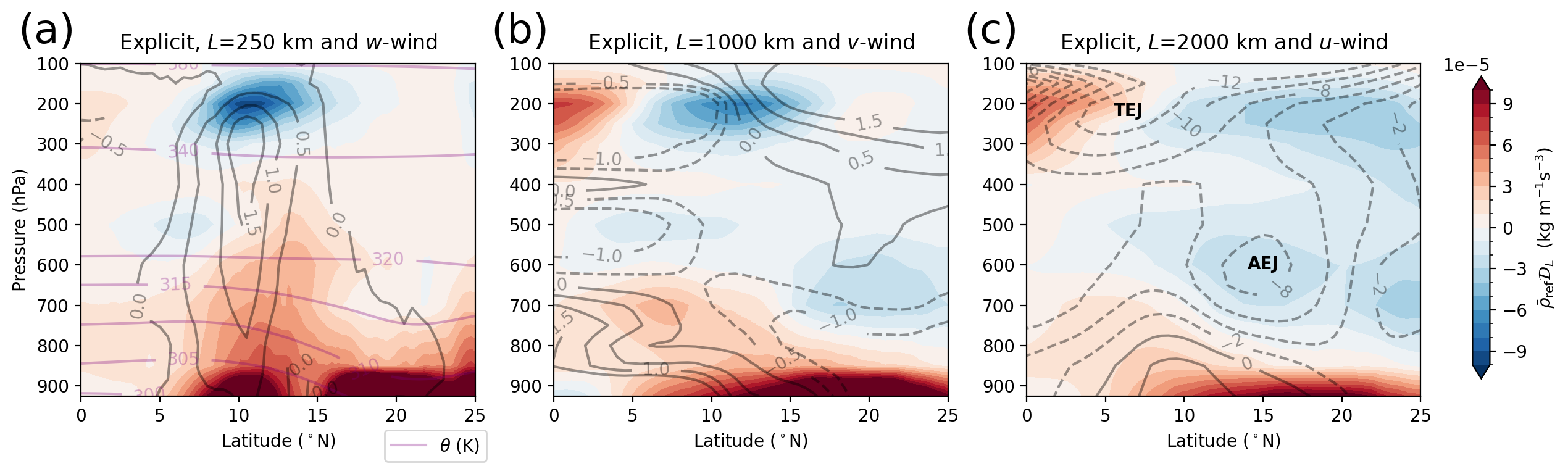}
    \includegraphics[width=\textwidth]{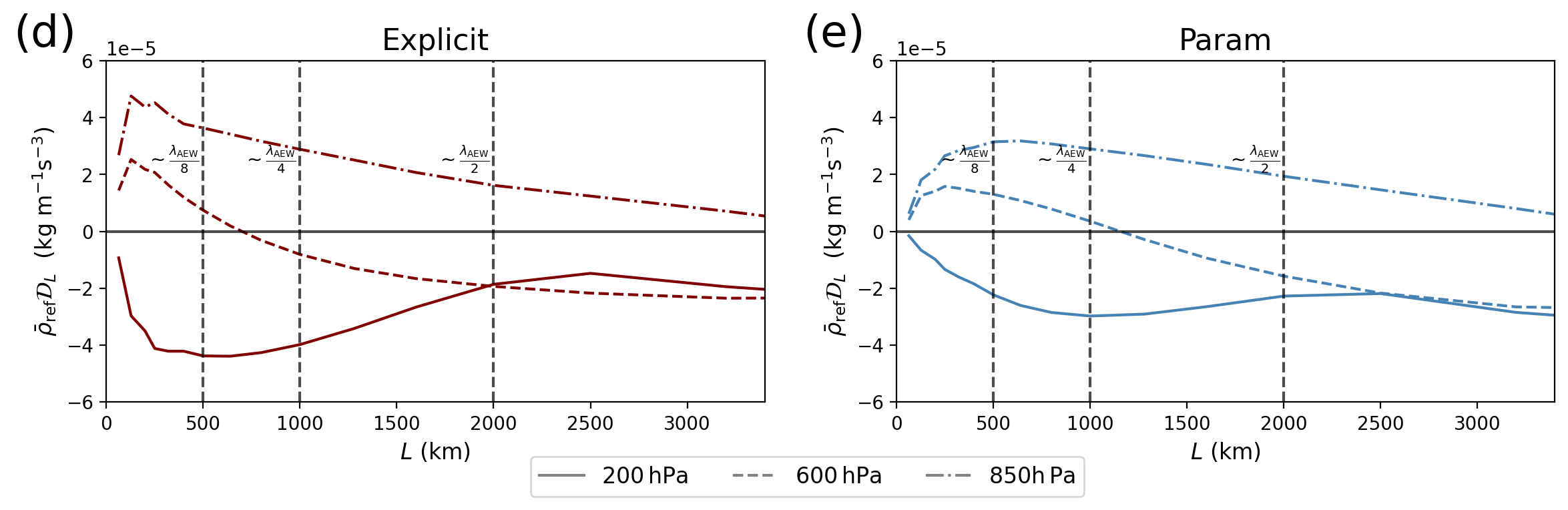}
    \caption{\textit{(a)} Full study--domain zonal mean mass--weighted $\iset$ at $L$=250\,km (shading) and vertical velocity $w$ (grey contours, units cm s$^{-1}$) for Explicit, calculated from -50--30$^\circ$E. Purple contours show potential temperature $\theta$ (K). Repeated for \textit{(b)} $L$=1000\,km and meridional velocity $v$ (units m\,s$^{-1}$) and \textit{(c)} $L$=2000\,km and zonal velocity $u$ (units m\,s$^{-1}$. The location of the TEJ and AEJ is labelled with respect to the zonal velocity contours. \textit{(d)} Mean $\rhoref\iset$ in Explicit at all scales $L$, plotted at 200\,hPa (solid curve), 600\,hPa (dashed) and 850\,hPa (dot--dash). Spatial mean taken over restricted meridional slice of 5--20$^\circ$N. Vertical dashed lines represent approximate scales $\lambda_{\rm AEW}/8$, $\lambda_{\rm AEW}/4$ and $\lambda_{\rm AEW}/2$ assuming an AEW wavelength $\lambda$ of 4000\,km. \textit{(d)} Repeated for Param.}
    \label{fig:ISETspectra}
\end{figure}

MCSs are a primary feature of West Africa's meteorology, and thus the behaviour of storm--centred energy transfers should scale up to full regional distributions of $\iset$ across scales. A first, coarse indication that this is indeed the case comes from simple mean zonal profiles of the interscale energy transfer in Explicit across the full study domain.  Figure~\ref{fig:ISETspectra}a shows that for $L$=250\,km, the characteristic MCS dipole pattern inprints upon the zonal mean in the Sahel region (9$^\circ$--19$^\circ$N), the climatological  maximum for West African MCS activity, and aligns with the mean ascending branch of the local overturning circulation (black contours). In Param (Fig.~S\SuppRef{}a) this upscale signal is weaker, reflecting a broader region of weaker ascent, and the dominant feature is instead centred on the mid--level AEJ. Both simulations show northwards--strengthening downscale energy transfer in the boundary layer; while there is likely a contribution from cold pool ventilation of the Sahara, comparable values between the simulations suggest processes such as strong turbulent mixing in the desert's deep boundary layer are more likely the primary driver.

\subsection{AEW length scales}
\label{sec:aew_length_scales}

AEWs are synoptic--scale modes of variability and thus associated with characteristic length--scales an order-of--magnitude larger than for MCSs. Increasing $L$ to 1000\,km (Fig.~\ref{fig:ISETspectra}b) and 2000\,km (Fig.~\ref{fig:ISETspectra}c) in the Explicit zonal means, we see the pattern of $\iset$ transition away from the strong MCS dipole and towards a weaker, broader upscale pattern at 2000\,km. At this scale, a local mid--level peak in $\iset$ at 15$^\circ$N aligns with the AEJ, shown by a local maximum in the easterly winds at 600\,hPa (contours), consistent with previous work showing that the jet plays a key role in communicating large--scale disturbances into the wider Atlantic basin \cite{Pante2019resolving}. Intriguingly, uniform areas of downscale energy transfer below 700\,hPa and south of 15$^\circ$N align with the prevailing monsoon flow, shown by the positive meridional velocity contours in Fig.~\ref{fig:ISETspectra}b; this likely reflects continuous alignment of small--scale eddies by the vertical shear in large--scale flow. This panel also continues to show strong 200\,hPa upscale energy transfer at $L$=1000\,km; the signal is weaker than at $L$=250\,km, but collocated and broader, showing that the impact of MCS convection on interscale energy transfer within the mean circulation continues to scales far larger than the systems themselves.

It is illustrative to form a complete background spectrum for the interscale energy transfer by considering how regional mean values of $\iset$ in Explicit vary with scale $L$ through the vertical profile. In Fig.~\ref{fig:ISETspectra}d we take the meridional average between 5$^\circ$--20$^\circ$N to isolate signals relevant for the southern AEW track, and represent the vertical profile with 200, 600 and 850\,hPa values. At all levels the spectrum peaks below 1000\,km, towards the MCS scale range but at $L>$500\,km. Mean energy transfer at 200\,hPa (solid) is always upscale (negative values), with a positive slope for large $L$, whereas at 850\,hPa (dot--dashed) the energy transfer is downscale (positive) for all $L$. Note that for results between $L$=2000\,km and $L$=5000\,km (the maximum calculated scale) we begin to extend beyond the latitudinal width of our AEW window and study domain; extended--domain spatial means are shown in Fig.~S2. An equivalent spectrum for Param is shown in Fig.~\ref{fig:ISETspectra}e: the qualitative shapes of the spectra are similar to Explicit, but the magnitude of $\iset$ is reduced and peaks are shifted to larger scales $L$.

As noted in the Introduction, the observed wavelength range for southern track AEWs is 2000--5000\,km, depending on both system and identification method \cite{Nunez2026review}. Let us therefore select a single representative AEW wavelength $\lambda_{\rm AEW}$=4000\,km, and consider mean $\iset$ at binomial factors of this scale. We then find that in Explicit, 200\,hPa upscale energy transfer (solid line) peaks at $\sim\lambda_{\rm AEW}/6$, whereas in Param the peak is closer to $\sim\lambda_{\rm AEW}/3$. Previous studies of interscale energy transfer in 2D isotropic turbulence have found that upscale transfer embedded in a scale $\lambda$ peaks between $\lambda/8$ and $\lambda/2$ \cite{chen2006,xiao2009}. This was recently found to hold for simulated 3D equatorial waves \cite{Mckinnon2025representation}, and indeed here we find similar results. We also find that the scales forcing the upscale transfer are smaller in a convection--permitting model, suggesting increased coupling to deep convection. 

At 600\,hPa, roughly the level of southern track AEW troughs, the $\iset$ peak in Explicit (Fig~\ref{fig:ISETspectra}d) is below $\lambda_{\rm AEW}/8$ and the slope then negative, with a change of sign at 700\,km and thus upscale energy transfer for larger scales. This change in sign aligns with the peak upscale transfer scale at 200\,hPa; we find a similar correlation in Param. There is an intriguing similarity between these spectra and simple two--layer models of geostrophic turbulence, where upscale transfer by mid--level barotropic modes occurs for scales larger than the Rossby radius of deformation $L_R$, due to kinetic energy transfer from an upper--level, downscale baroclinic mode \cite{Salmon1998lectures}. From standard formulae for the tropical atmosphere \cite{Raymond2015balanced,Dhavale2025structure}, we estimate that for AEWs, $L_R\sim$525\,km. Similarly, the closely--related Rhines scale (computed following \citeA{Dritschel2008multiple}) is estimated as $\sim$600\,km. See Supplementary Information for details. The 600\,hPa spectrum thus qualitatively aligns with geostrophic turbulence, which relates to interscale energy transfer between purely rotational flows. However, while the 200\,hPa spectrum peaks in Figs.~\ref{fig:ISETspectra}d,e are closely aligned with the 600\,hPa sign change, the spectrum is upscale for all $L$, with a strong forcing (i.e. upscale $\iset$ and negative slope) from small--$L$ values evident. This is markedly different to the large--$L$, downscale forcing in geostrophic turbulence models, and reflects the fact that the West African atmosphere is not purely rotational: there are strong divergent outflows caused by convective activity.
\begin{figure}[h!]
    \centering
    \includegraphics[width=\textwidth]{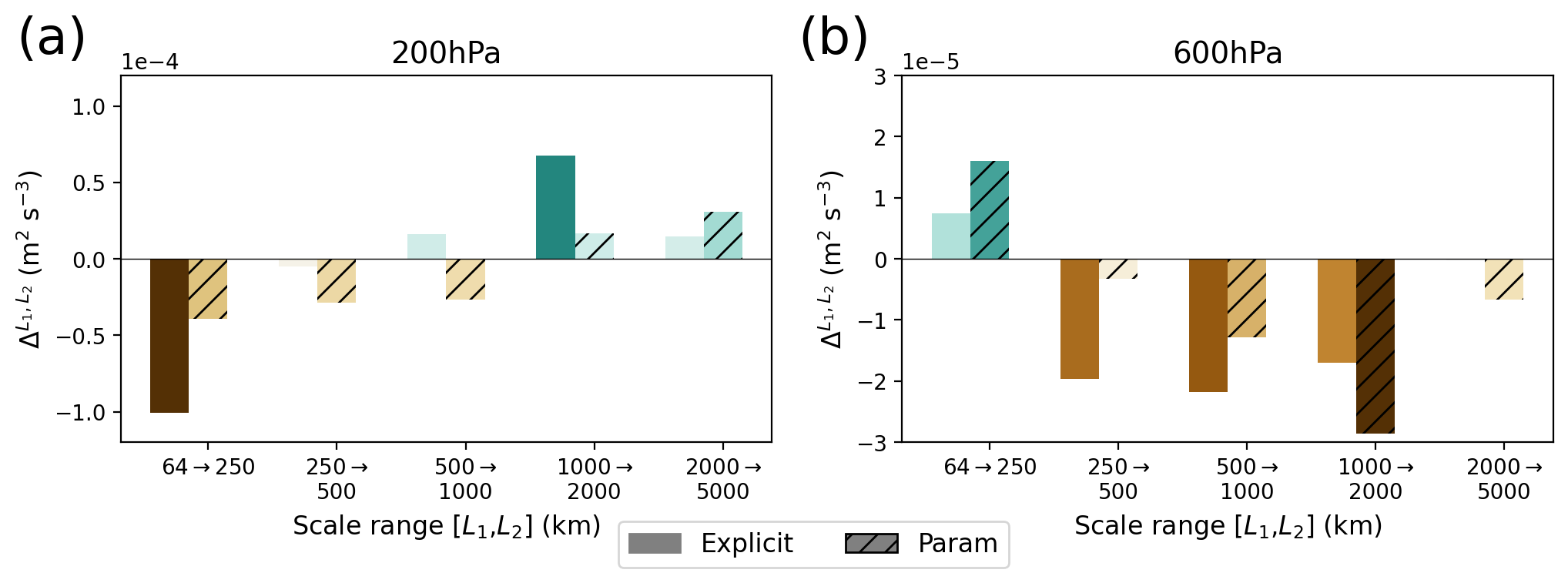}
    \includegraphics[width=\textwidth]{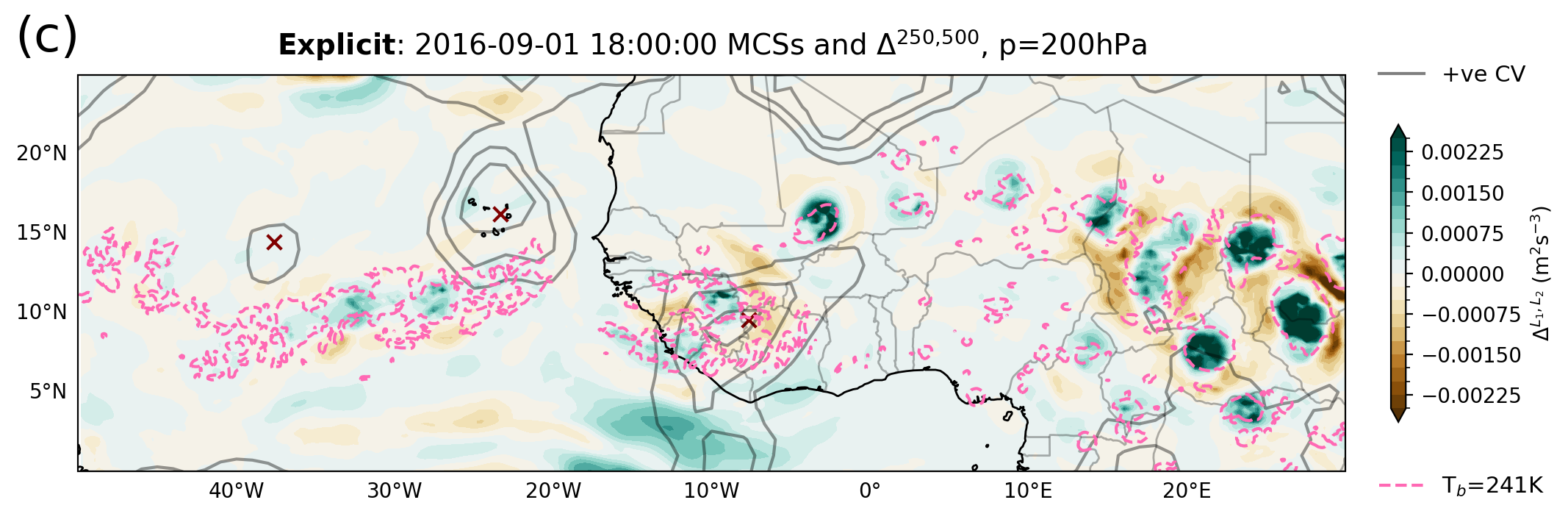}
    \includegraphics[width=\textwidth]{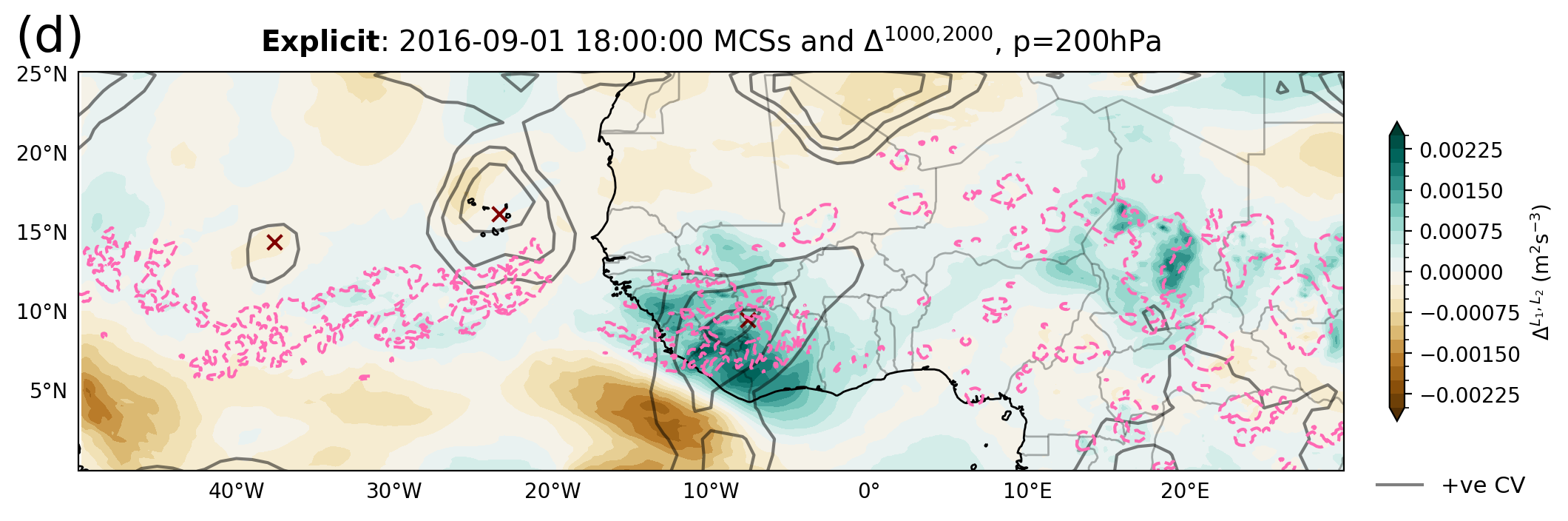}
    \caption{Full study domain mean convergence of interscale energy transfer $\Delta^{L_1,L_2}$ across scale bands at \textit{(a)} 200\,hPa and \textit{(b)} 600\,hPa, in Explicit (solid bars) and Param (hatched). Axis ranges differ between panels since values are not mass--weighted. \textit{(c)} Instantaneous snapshot of $\Delta^{L_1,L_2}$ for the scale band 250$\to$500\,km in Explicit at 18 UTC, 01/09/2016. Pink dashed lines show the $T_b$=241K contour, while grey contours show positive curvature vorticity at intervals of 2$\times$10$^{-6}$s$^{-1}$, and crosses tracked AEW troughs. \textit{(d)} Repeated in Explicit for scale band 1000$\to$2000\,km.}
    \label{fig:regionalISETbands}
\end{figure}

To explore the relationship between the interscale energy transfer spectrum and large--scale flow further, we compute its convergence across a scale band $[L_1, L_2]$, defined as
\begin{equation}
   \Delta^{L_1,L_2} = \mathcal{D}_{L_2}(\mathbf{u}) - \mathcal{D}_{L_1}(\mathbf{u}),\, \qquad L_2>L_1\,.
\end{equation}
This measure crudely approximates the slopes in Figs.~\ref{fig:ISETspectra}d,e and is thus essentially a local version of the spectral energy transfer function in traditional Fourier analysis of turbulence. $\Delta^{L_1,L_2}$ measures whether there is net energy input into, or extracted from, the scale band $[L_1,L_2]$. Figure~\ref{fig:regionalISETbands}a shows that at 200\,hPa in Explicit (solid), on average there is strong divergence of energy transfer from sub--MCS scales (below 250\,km), and peak convergence in the 1000$\to$2000\,km scale band that coincides with AEW trough scales. From the solid curve in Fig.~\ref{fig:ISETspectra}d, we see that the divergence reflects the net upscale transfer of energy out of the scale band: mean upscale $\iset$ is stronger at 250\,km than 64\,km. Scale bands beyond the $\sim$700\,km peak are then convergent since the upscale magnitude of $\iset$ decreases at larger $L$. In Param (hatched) the overall pattern is similar, but shifted to larger scales, with only weak convergence in the AEW trough--scale range.

Plotting a single spatial Explicit snapshot of the 200\,hPa 250$\to$500\,km scale band (Fig.~\ref{fig:regionalISETbands}c), we see that spatial regions of convergent energy transfer into this scale band are spatially localised and collocated with MCSs, i.e. expansive contours of cold cloud temperatures (pink contours). The snapshot is representative of the MCS composite mean (Fig.~S\SuppRef{}a). The upscale convergence of energy to meso--$\alpha$ scales is thus driven by MCSs, where deep convection generates strong lower--$L$ upscale energy transfer. The confined upscale convergence within MCS anvils is also stronger than surrounding downscale $\bandiset$; negative $\bandiset$ halos correlate with subsidence regions around storms where the 0.5$^\circ$ flow is convergent (not shown). At larger scales, e.g. the 1000$\to$2000\,km band in which mean $\bandiset$ peaks, the convergence of energy is no longer strictly localised to MCS anvils (Fig.~\ref{fig:regionalISETbands}d), being far more dispersed than in the lower--$L$ scale band. Convergent $\bandiset$ remains loosely tied to regions of MCS convection, with the primary positive arc over West Africa following the distribution of convection, while MCS--composite $\bandiset$ at these scales shows a local enhancement of a uniformly positive background (Fig.~S5b). The 1000$\to$2000\,km signals no longer correlate with the locally divergent flows, suggesting interscale energy transfer at AEW scales results from the interaction of embedded divergent flows with the ambient synoptic rotational flow (e.g.~\citeA{kouhen2024}).

In Param, we do not find spatial localisation of convergence of interscale energy tranfer within MCS anvils. This can be seen both in equivalent representative snapshots (Fig.~S4d,e) and MCS--centred composites (Fig.~S5c,d). Parameterising deep convection thus fundamentally changes the contribution of MCSs to regional energetics, and in particular shifts the convergence of energy from the upscale energy cascade to much larger scales, perhaps reflecting weaker divergent--rotational flow interactions. The largest--magnitude $\bandiset$ values in Param at 600\,hPa are also shifted to larger $L$ values than in Explicit (Fig.~\ref{fig:regionalISETbands}). Referring to the dashed curves in Fig.~\ref{fig:ISETspectra}d,e, we see that convergence at the smallest spatial scales at 600\,hPa is in fact downscale from the spectral peak. Both simulations show divergent energy transfer across bands greater than 250\,km, which is upscale beyond 1000\,km, i.e. for $L$ reaching AEW scales. These results suggest AEWs themselves cause mid--level upscale energy transfer that peaks at around $\lambda_{\rm AEW}$. However in Explicit, the upscale transfer embedded within the waves (and thus acting as a potential driver) appears to stem from MCS scales at 200\,hPa. It is the details of this question to which we turn in the next section.


\section{Upscale energy transfer to African Easterly Waves}
\label{sec:mcs}

\begin{figure}[t]
    \centering
    \includegraphics[width=\textwidth]{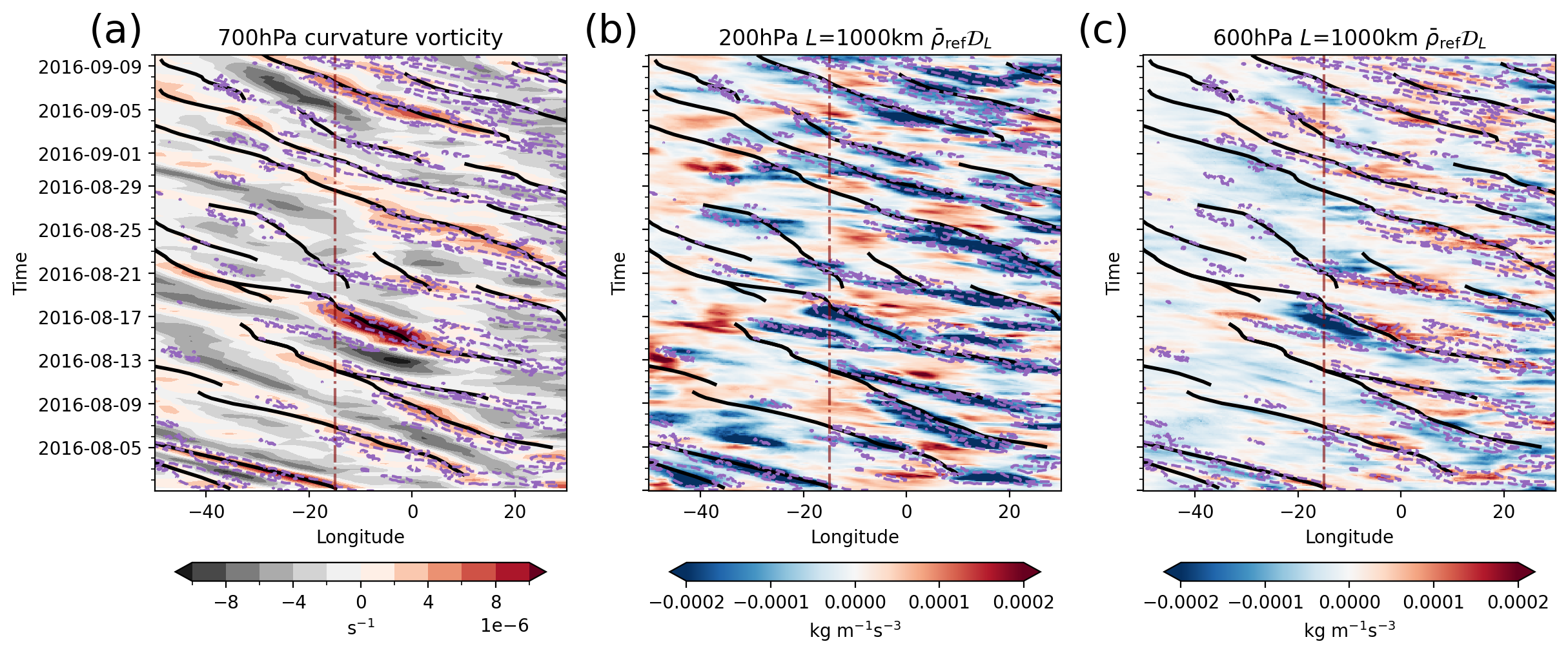}
    \includegraphics[width=0.9\textwidth]{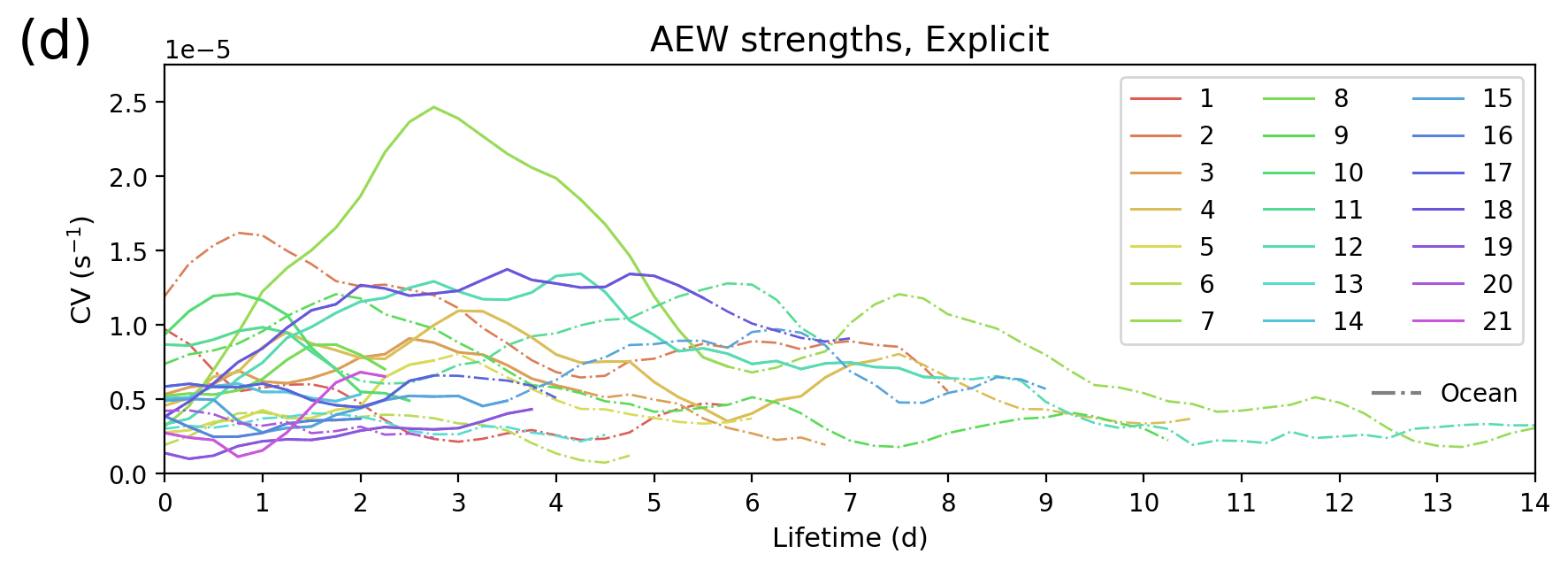}
    \caption{\textit{(a)} Hovm\"{o}ller of curvature vorticity at 700\,hPa in Explicit, with meridional average taken from 5$^\circ$--20$^\circ$N. Overlain are tracked AEW trough longitudes (black lines) from QTrack, and regions of cloud with $T_b$ $<$230K (dashed purple lines; same meridional slice). Repeated for $L$=1000\,km $\rhoref\iset$ at \textit{(b)} 200\,hPa, and \textit{(c)} 600\,hPa. \textit{(d)} Grouped plots of individual evolution of AEW strength for each tracked system in Explicit post initiation. Line plots have been smoothed to a 1--day rolling mean. Systems are labelled in temporal order, with lines dashed for ocean--phase AEWs (west of 15$^\circ$W).}
    \label{fig:hovmoellers}
\end{figure}

We now focus directly on the relationship between AEWs and interscale energy transfer, only considering results from Explicit in this Section. Figure~\ref{fig:hovmoellers}a shows a Hovm\"{o}ller diagram of 1$^\circ$ curvature vorticity for the full simulation, using a meridional average centred on the Sahel to capture AEW activity. Time is moving forwards going up the $y$--axis, and the diagram covers continental West Africa and the East Atlantic, with the coast delineated by a dashed line at 15$^\circ$W. Objective AEW tracks from QTrack (black lines) show good alignment with bands of positive curvature vorticity (CV). There are 21 ostensibly independent tracks, with the Hovm\"{o}ller showing several are likely parts of the same weak systems. Cold cloud contours ($T_b<$230K, purple) are also overlain on the Hovm\"{o}ller to represent the location of MCS activity. The documented observed pattern of MCSs (cold cloud cover) leading AEW troughs (positive CV regions) is well simulated by Explicit. This simulation also appears to capture the observed discrepancy in MCS versus AEW speeds, with cloud bands often showing faster velocities around the coast in particular.

Figures \ref{fig:hovmoellers}b and~\ref{fig:hovmoellers}c repeat the same Hovm\"{o}ller, but with the shading changed to show 200\,hPa and 600\,hPa $L$=1000\,km mass--weighted $\iset$, respectively. These fields are chosen such that Fig.~\ref{fig:hovmoellers}b can be considered to span anvil--level ``MCS physics'', and Fig.~\ref{fig:hovmoellers}c to represent mid--level ``AEW physics''. Consistent with Fig.~\ref{fig:mcsISETcomps}, the cold cloud anvil regions are accompanied by tightly--collocated bands of strong upscale energy transfer that lie ahead of, or in, the AEW troughs. Away from the cold cloud regions there are some weaker upscale regions, primarily over the ocean, but overall a primarily weak, dispersed downscale background. At 600\,hPa (Fig.~\ref{fig:hovmoellers}c) values of $\iset$ are weaker (common scale bars) but the visible features are broader than at 200\,hPa and no longer tied to the anvil bands. The relationship between AEW troughs and $\iset$ is more diverse than for the convective systems. For many tracks, the sign of $\iset$ varies along the track; the magnitude of the energy transfer is notably stronger over land than the ocean. We generally find that large regions of downscale transfer follow the passage of wave troughs, and in many cases upscale transfers lead the troughs. This is particularly the case for the very strong curvature vorticity region evident around 10$^\circ$W on 16$^{\rm th}$ August, which follows upscale energy transfer at both 200\,hPa and 600\,hPa. We shall return to this particular case shortly.

\begin{figure}[t]
    \centering
    \includegraphics[width=\textwidth]{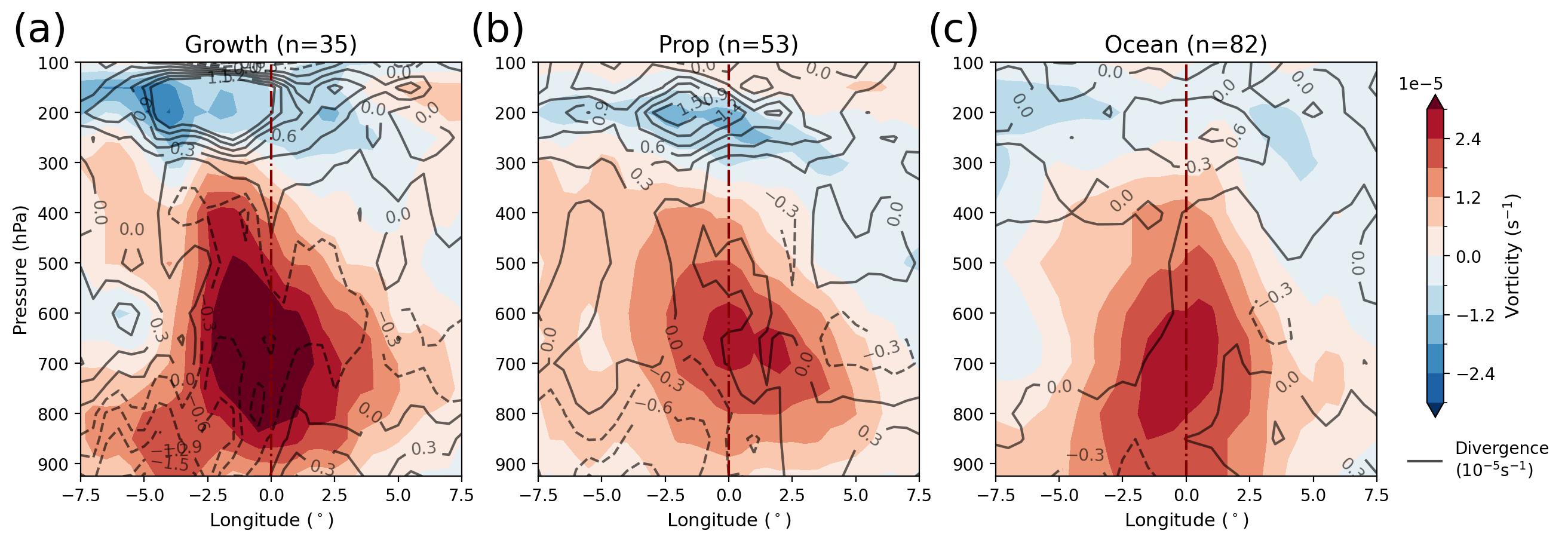}
    \includegraphics[width=\textwidth]{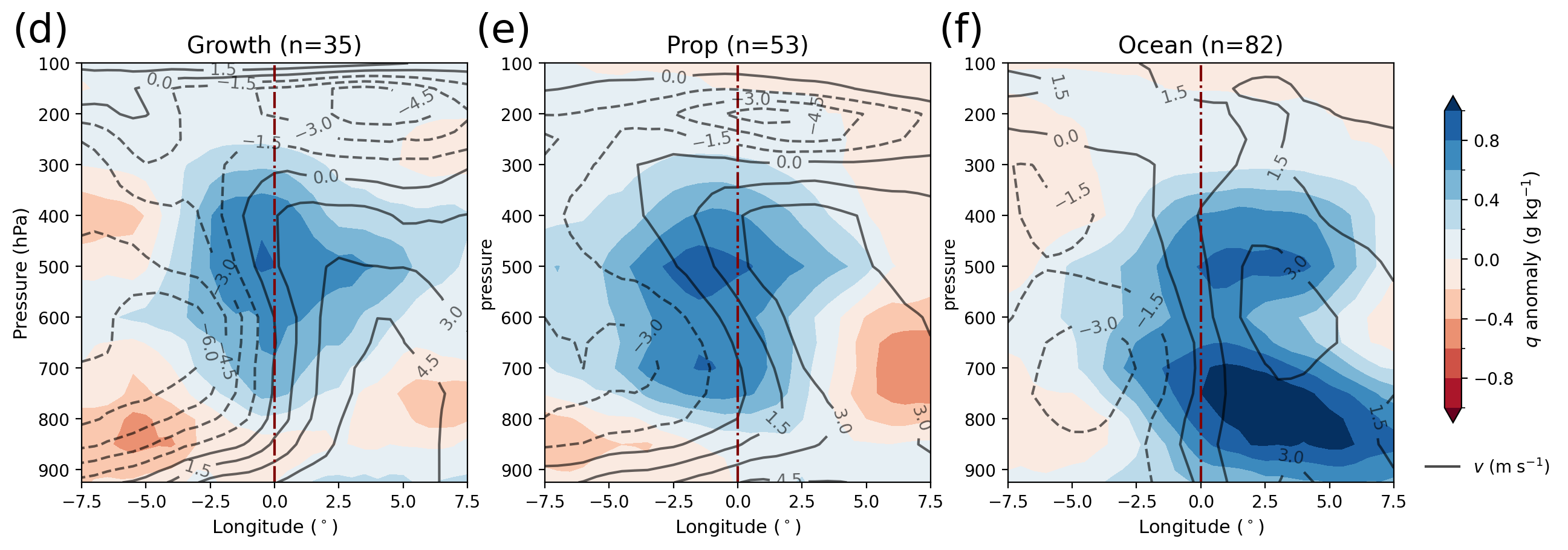}
    \caption{AEW trough--centred composites in Explicit for growth, propagation and ocean life--phases. Shading in panels \textit{(a--c)} show relative vorticity, with contours displaying horizontal divergence. Panels \textit{(d--f)} show specific humidity $q$ anomalies (shading) and meridional winds $v$ (contours). In all panels, latitudinal averages are taken over central 5$^\circ$ slice. Values in parentheses show the number of tracked troughs in each AEW phase.}
    \label{fig:aew_comps}
\end{figure}

The track overlays in Figs.~\ref{fig:hovmoellers}a---c hide the differing evolution of AEW troughs in Explicit. Figure~\ref{fig:hovmoellers}d shows the time evolution of AEW strength identified by QTrack as trough CV, for all 21 systems. Lines are solid when the track lies over land. We find that even in this 40 day simulation, there is a marked variability in AEW strengths and vorticity amplification rates. AEW strength is not necessarily a monotonic function of lifetime; systems can re--intensify after decaying, pointing to potential influences from interactions with their environment. Plotting composite vertical zonal sections of the wind field and moisture anomalies centred on trough locations (Fig.~\ref{fig:aew_comps}) confirms that the AEWs in Explicit exhibit canonical structures of observed southern--track AEWs \cite{Nunez2026review}: this is not our focus here. Rather, we are interested in the division into different phases of the AEW lifecycle. Note that the composites cover a much larger spatial window than those in Fig.~\ref{fig:mcsISETcomps}.

Based on Fig.~\ref{fig:hovmoellers}d, we classify AEW troughs based on whether they are in a ``growth'', ``propagation'' or ``ocean'' phase, and take composites for each group separately. Growing AEWs are those east of 15$^\circ$W where dCV/d$t>$ 10$^{-5}$\,s$^{-1}$d$^{-1}$. Propagating AEWs are then all other systems east of 15$^\circ$W, and oceanic AEWs are all systems west of 15$^\circ$W, i.e. offshore. AEWs of course still propagate westwards in the growth and ocean phases. As expected, relative vorticity and divergence fields are strongest in the growth phase (Fig.~\ref{fig:aew_comps}a). For the growth and propagation (Fig.~\ref{fig:aew_comps}b) phases, the divergence field (contours, top row) shows a characteristic MCS dipole signature leading the wave trough that is not apparent in the ocean phase (Fig.~\ref{fig:aew_comps}c). In all phases there is a mid--level moisture anomaly centred around 500\,hPa, however this is weakest in the growth phase (Fig.~\ref{fig:aew_comps}d). The lifecycle phases are distinguished by very different tilts in the meridional wind (contours, bottom row): in the growth phase, there is an eastward (upshear) tilt beneath the 600\,hPa AEJ level, and westward (downshear) tilt above the jet level. This is consistent with baroclinic growth mechanisms, which contribute to AEW development and amplification. In the propagating phase (Fig.~\ref{fig:aew_comps}e) the tilt is entirely downshear except in the boundary layer, while over the ocean (Fig.~\ref{fig:aew_comps}f) there is no tilt, due to the absence of a baroclinic jet. Similar meridional wind structures have been documented by many authors \cite{Kiladis2006three,Yang2018linking,Nunez2026review}.

\begin{figure}[h!]
    \centering
    \includegraphics[width=\textwidth]{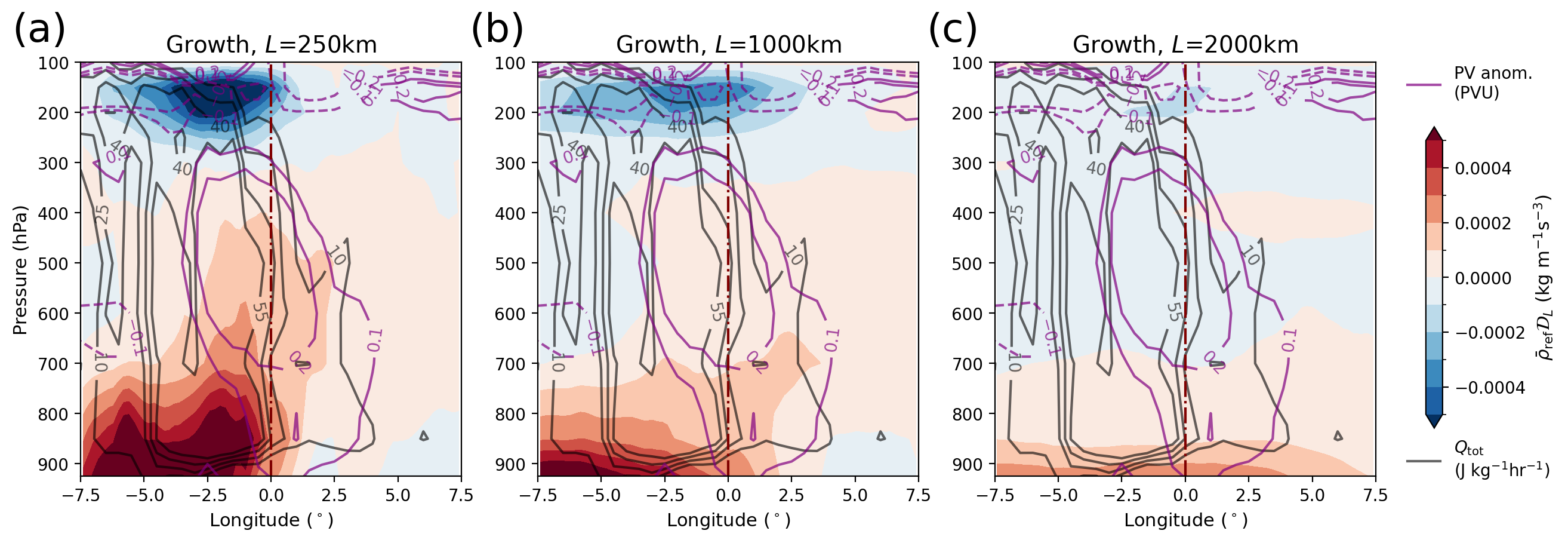}
    \includegraphics[width=\textwidth]{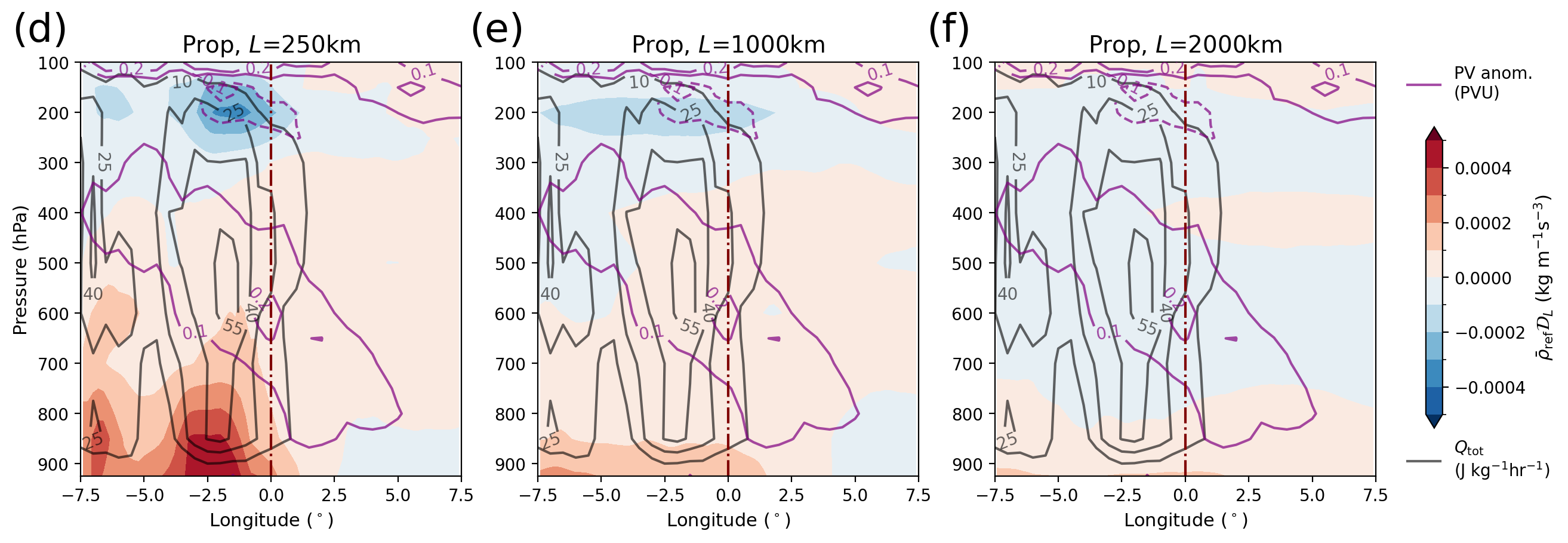}
    \includegraphics[width=\textwidth]{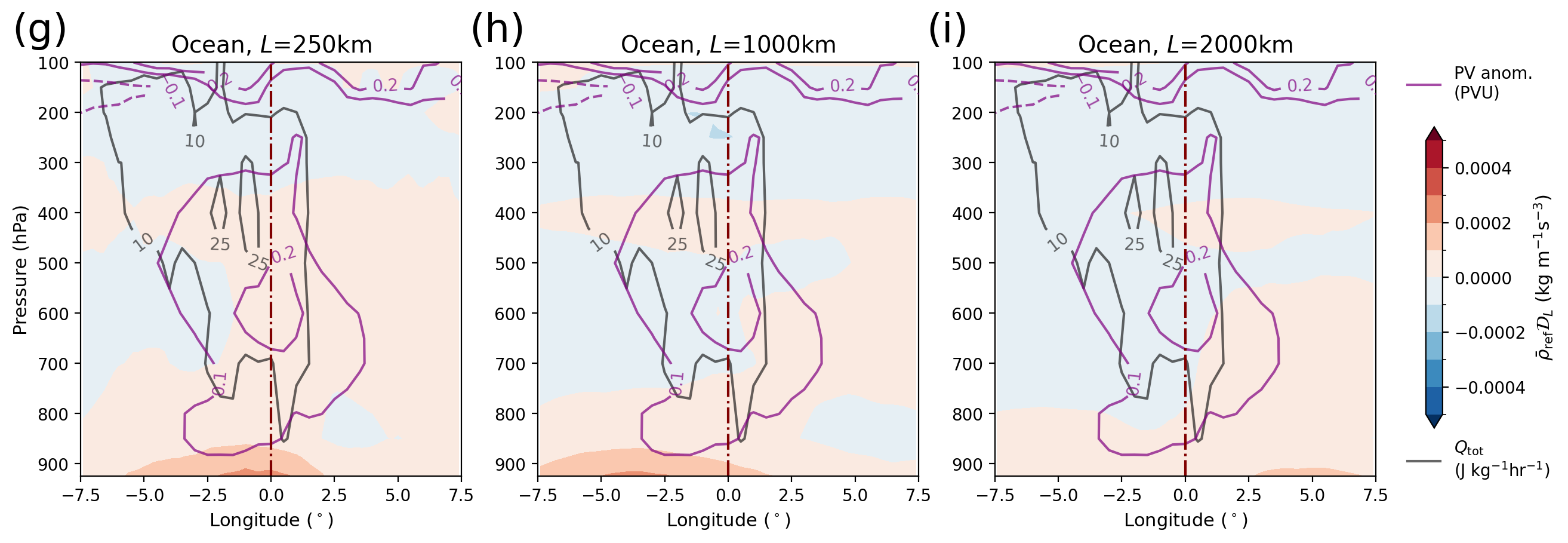}
    \caption{AEW track--centred composite vertical sections in Explicit of $\rhoref\iset$, grouped by life--phase (rows) and scale $L$ (columns). Top row shows growth--phase mean interscale energy transfer at \textit{(a)} $L$=250km, \textit{(b)} $L$=1000km and \textit{(c)} $L$=2000km. Second row, panels \textit{(d--f)}, repeated for propagation phase; and third row, panels \textit{(g--i)}, repeated for oceanic AEWs. In all rows, black contours indicate composite phase $\Q$, and purple contours show potential vorticity anomalies. All latitudinal means taken over central 5$^\circ$ slice, and scales are common between all panels.}
    \label{fig:aewISETcomps}
\end{figure}

In addition to baroclinic growth signatures, the composites hint at a role of upscale energy transfer in AEW growth: growing wave troughs show strong MCS divergence signatures, which from Fig.~\ref{fig:mcsISETcomps} we know mirrors their $\iset$ signatures. Figure~\ref{fig:aewISETcomps} therefore shows equivalent vertical composites to Fig.~\ref{fig:aew_comps} for $\iset$ (shading), $\Q$ (black contours) and PV anomalies (purple contours). AEW phases are grouped by row, with each column sharing a common scale $L$. PV and $\Q$ contours are shared by panels on the same row.

Beginning with the first row of panels for growing AEWs, we find consistent dipole patterns of $\iset$ for all scales $L$, with stronger values at smaller scales. At $L$=250\,km (Fig.~\ref{fig:aewISETcomps}a) the dipole pattern is very similar to that for MCSs in strongly sheared environments shown in Fig.~\ref{fig:mcsISETcomps}, and, consistent with the divergence pattern in Fig.~\ref{fig:aew_comps}a, lies west of the composite centre. The growth--phase AEW composite $\iset$ at MCS scales is therefore indeed dominated by signatures of MCS interscale energy transfer. This continues to hold at $L$=1000\,km (Fig.~\ref{fig:aewISETcomps}b). In contrast, for propagating waves there is only a very dilute $L$=1000\,km MCS dipole signature leading the wave trough (Fig.~\ref{fig:aewISETcomps}e). Such a signal is clearer at $L$=250\,km (Fig.~\ref{fig:aewISETcomps}d), but significantly weaker than for growing waves. For both phases, weak mid--level upscale transfer is apparent at $L$=2000\,km. For growing AEWs (Fig.~\ref{fig:aewISETcomps}c), peak $L$=2000\,km upscale transfer remains at 200\,hPa but lies at the composite axis, rather than leading the trough, indicating that upscale transfer at this scale is primarily associated with the stratiform, rather than convective, dynamics of MCSs. We find no MCS $\iset$ signature for oceanic AEWs (Figs~\ref{fig:aewISETcomps}g--i), with relatively uniform patterns across the scale range reflecting the general weakness of the oceanic trough population in Explicit (Fig.~\ref{fig:hovmoellers}d). 

Returning to Fig.~\ref{fig:aewISETcomps}a, we see that the vertical axis of the $L$=250\,km $\iset$ dipole is aligned with that of the $\Q$ heating contours at $x$=\,-2.5$^\circ$, as found in Fig.~\ref{fig:mcsISETcomps} for MCSs. The strong upscale transfer region is also located directly above the western extent of a downshear--tilted mesoscale PV maximum (purple contours), and coincident with upper--level negative (dashed) PV anomalies of similar magnitude. Such a vertical dipole is characteristic of diabatic PV generation and expected for moist baroclinic waves \cite{de2010baroclinic}. Meanwhile the relative location of the mid--level positive PV anomaly is consistent with PV tendencies arising from sharp vertical temperature gradients caused by MCS stratiform heating, which are the primary contributor to AEW PV growth tendencies \cite{Johnson2020potential,Russell2020potential}. We find that the composite absolute PV is positive at all levels (Fig.~S\SuppRef{}). The PV dipole formation, and upper--tropospheric upscale transfer of energy, are therefore both consistent with upward diabatic mass transport across isentropic layers and dilution of PV in the outflow region \cite{Methven2015potential,Saffin2021circulation}. This mechanism generates negative PV anomalies, but cannot change the absolute sign of the PV field \cite{Harvey2020diabatic}. Such mechanisms, identified in the context of midlatitude warm--conveyor belt outflows but also applied to tropical waves \cite{Hardy2023examining}, provide a formal understanding of how sub--convective scale eddies can grow when embedded within a larger--scale divergent flow, and thus transfer energy upscale to synoptic scales.

The composite PV anomalies and diabatic heating for growing AEWs are notably stronger than in the propagation or oceanic phases (rows 2 and 3 in Fig.~\ref{fig:aewISETcomps}). These phases also show very few negative PV anomalies, consistent with the reduced $\iset$ dipole signature from leading MCSs, while the positive PV anomaly for propagating AEWs (e.g.~\ref{fig:aewISETcomps}d) shows a stronger downshear tilt than for growing waves. This is consistent with reduced low--level PV generation by MCS convective updrafts \cite{Russell2020potential}. In the propagating and oceanic phases of AEW lifecycles we therefore find a reduced role of diabatic processes compared to during wave growth. The contrasting $\iset$ signatures to the growth phase moreover point to a role of upscale energy transfer from MCSs in AEW amplification. To examine whether the $\iset$ signals actually contribute to wave growth, rather than simply correlating, we examine a case study of AEW growth and decay.

\subsection{Case study of a strengthening AEW}

\begin{figure}[h!]
    \centering
    \includegraphics[width=0.495\textwidth]{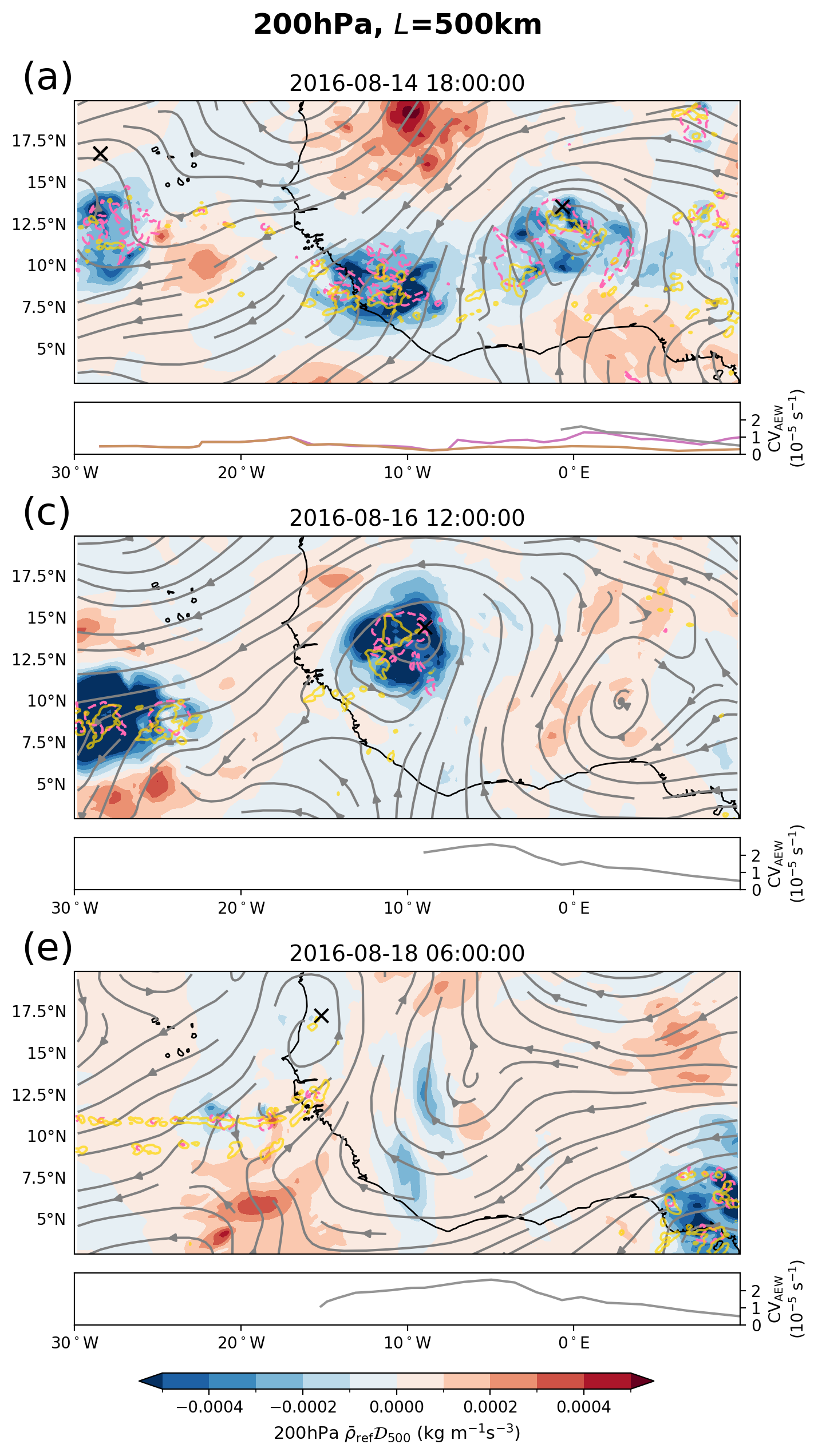}
    \includegraphics[width=0.495\textwidth]{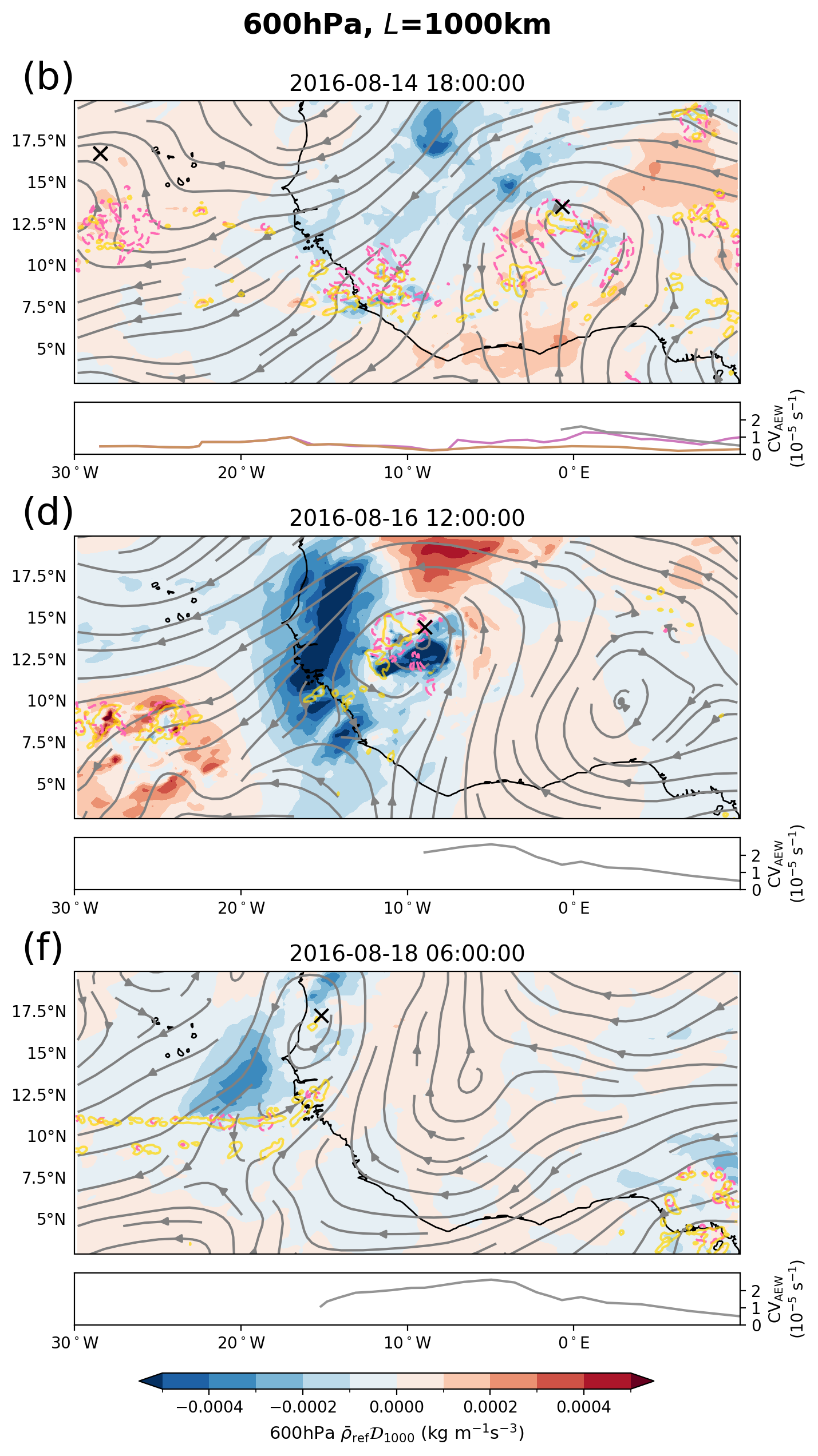}
    \includegraphics[width=0.8\textwidth]{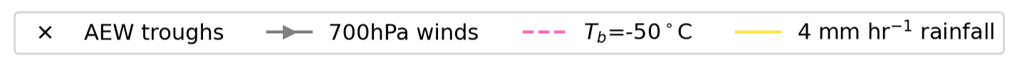}
    \caption{Case study of strongest AEW (system 7) in Explicit. Rows correspond to common timesteps, columns common interscale energy transfer scale $L$. \textit{(a)} 200\,hPa $L$=500\,km $\rhoref\iset$ (shading), 700\,hPa winds (streamlines), and tracked AEW troughs (crosses) over West Africa at 18 UTC on 14/08/2016. Contours show regions of $T_b$ $<$-50$^\circ$C (pink, = MCSs) and rainfall $>$ 4\,mm\,hr$^{-1}$ (gold, = convective rain). Inset below map plots evolution of AEW curvature vorticity for tracked systems displayed in panel, on common longitudinal axis, showing that the AEW (grey) is strengthening. \textit{(b)} Shading shows 18 UTC 14/08/2016 $L$=1000\,km $\rhoref\iset$ at 600\,hPa, with all other quantities common to \textit{(a)}. \textit{(c--d)} Panels repeated for 12 UTC on 16/08/2016, when the AEW has reached maximum strength (insets). \textit{(e--f)} Panels repeated for 06 UTC on 18/08/2016, when the AEW trough is decaying (insets). Columns plotted on common colour scales.}
    \label{fig:case_study}
\end{figure}

Figure~\ref{fig:hovmoellers}d shows that the AEW population contains a system (wave 7) that lasts 8 days and achieves a curvature vorticity nearly double that of other troughs. This is the same system noted earlier as standing out in Hovm\"{o}ller plots around 10$^\circ$W on the 16$^{\rm th}$ August. Figure~\ref{fig:case_study} therefore shows maps of (mass--weighted) $\iset$ over West Africa centred around 16$^{\rm th}$ August (middle row), when the wave strength peaks (inset panels, grey). The left column plots ``MCS--scale'' ($L$=500\,km) 200\,hPa $\iset$, the right column ``AEW--scale'' ($L$=1000\,km) 600\,hPa $\iset$, while rows share common timestamps and large-scale winds (700\,hPa streamlines).

At 18\,UTC on 14$^{\rm th}$ August the top row of Fig.~\ref{fig:case_study} shows that the wave trough is centred at 0$^\circ$E, 12$^\circ$N (streamlines and track location) and is mid--way through its long growth phase (inset CV panels). At 200\,hPa (Fig.~\ref{fig:case_study}a) there are comparably--sized clusters of strong upscale energy transfer collocated within the wave trough and to its south--west. The sample time of 18 UTC is the diurnal peak of convective activity, and as expected we find MCSs (pink cold--cloud and gold heavy rainfall contours) embedded within the upscale regions. Crucially, numerous MCS anvil shields are clustered within the wave trough. A similarly structured cluster of MCS anvils and (stronger) upscale transfer is apparent on the Guinea coast, 10$^\circ$W of the AEW trough. At 600\,hPa (Fig.~\ref{fig:case_study}b) the 1000\,km $\iset$ is primarily downscale within the trough, but with some weak signatures of upscale transfer emerging around the strongest MCS activity.
\begin{figure}[t]
    \centering
    \includegraphics[width=\textwidth]{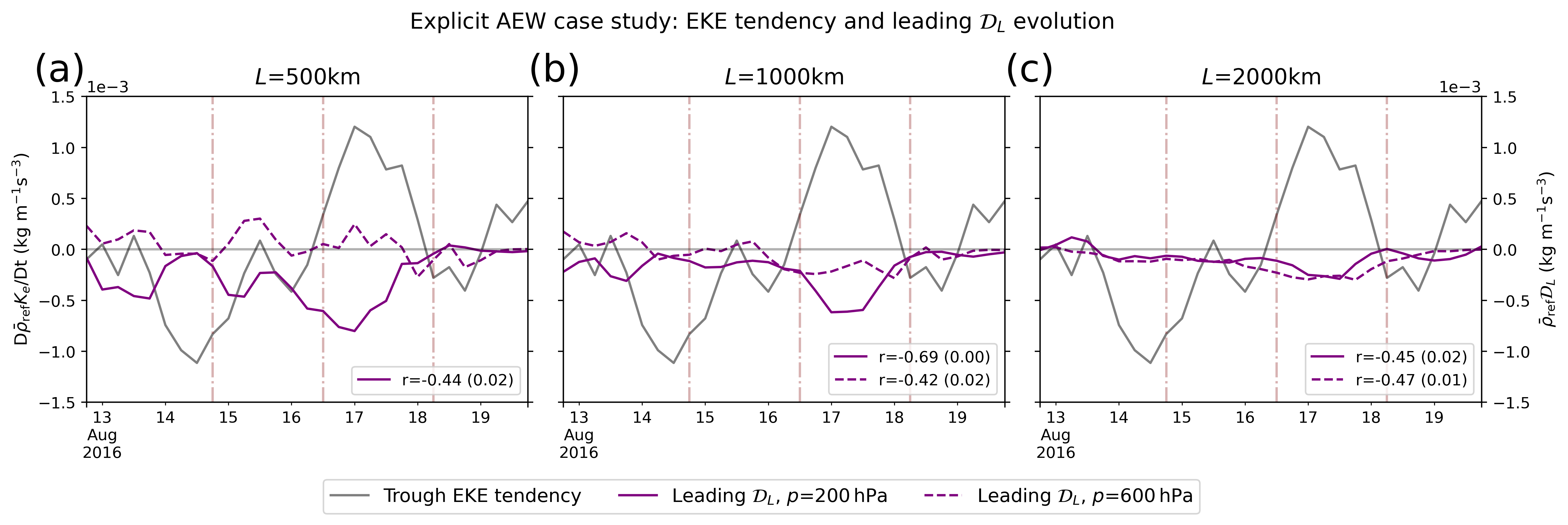}
    \caption{Evolution of trough mean (Lagrangian) EKE tendency (grey) for case study Explicit AEW (wave 7). Accompanied by leading 200\,hPa (solid purple) and 600\,hPa (dashed) $\iset$ computed at \textit{(a)} $L$=500\,km, \textit{(b)} $L$=1000\,km and \textit{(c)} $L$=2000\,km. All measures are mass--weighted and thus on common scales. Trough $\textrm{D}K_e/\textrm{D}t$ computed over 300\,km radius centred on tracked AEW trough; $\iset$ measures similar, but offset 2$^\circ$W. Statistically significant Pearson correlation values given in legends; vertical maroon lines indicate times of panels in Fig.~\ref{fig:case_study}.}
    \label{fig:case_energetics}
\end{figure}

After a further full diurnal cycle, the AEW trough reaches its maximum intensity. During this cycle, the MCSs embedded within the trough do not decay and reform, but rather aggregate to form a single, long--lived system. There is therefore a source of significant, coherent upper level upscale energy transfer (Fig.~\ref{fig:case_study}c) embedded directly within the wave trough. The consistent 200\,hPa upscale signal from MCS--scales and its impact on the AEW can be seen in Fig.~\ref{fig:case_energetics}, which tracks the trough--mean EKE tendency and leading $\iset$ evolution. Here EKE (Eq.~\ref{eqn:eke}) is calculated with respect to time means, which then satisfies a local budget previously applied to AEW dynamics by \citeA{Diaz2013energy}. Based on the composites in Figs.~\ref{fig:aew_comps} and~\ref{fig:aewISETcomps}, we calculate the AEW EKE tendency as the mean 750--600\,hPa (Lagrangian) tendency over a 300\,km radius footprint centred on the tracked wave trough. The $\iset$ contributions from a given level are calculated over the same size footprint, but from a centre 2$^\circ$W of the trough, reflecting the position of the MCS dipole (Fig.~\ref{fig:aewISETcomps}a,b). From these metrics, we indeed see that from 18 UTC on 14/08/2016, consistent upscale transfer at 200\,hPa, $L$=500\,km (solid purple, Fig.~\ref{fig:case_energetics}a) leads the wave trough, and its EKE tendency, until a peak at 00 UTC 17/08/2016. Over the AEW's lifecycle, there is a significant correlation of $r$=-0.44 between $L$=500\,km, 200\,hPa $\iset$ and the EKE tendency. There are no such equivalent signals at 600\,hPa (dashed).

This long--lasting forcing exerts an influence at the level of the trough, with upscale transfer at $L$=1000\,km found embedded within the trough at 600\,hPa by 16/08/2016 (Fig.~\ref{fig:case_study}d). From Fig.~\ref{fig:case_energetics}, we see that this lags the $L$=500\,km signal at 200\,hPa; the larger--$L$ signals at 200\,hPa are also upscale; and that the larger--$L$ upscale signals occur in--phase with the AEW EKE tendency. Correlations at both $L$=1000\,km and $L$=2000\,km to EKE tendency are significant at both atmospheric levels. Together, these results show energy from diabatic heating is transferred upscale from the embedded MCS to the EKE of the AEW trough, with this upscale transfer causing a positive EKE tendency. The magnitude of the maximum $L$=500\,km upscale transfer is two--thirds of the peak EKE tendency. The upscale energy transfer is of such a high magnitude in this case because of the extremely large areal extent of the MCS, embedded exactly within the wave trough. Numerous studies point to the crucial role of MCS area coverage within AEW troughs to AEW amplification in the context of tropical cyclogenesis \cite{Leppert2013relation,Price2015west,Nunez2020wave}: this case study supports this perspective, while we also remark that the case MCS also delivers the most extreme West African point--rainfall accumulations of the full simulation period.

It is important to note that the peak absolute and tendency EKE values lag the curvature vorticity, which is slightly decreasing at the EKE peak on 17/08/2016. Vorticity budget analysis (not shown) confirms that vortex stretching is the primary contributor to the CV amplification, providing complementary evidence that the AEW amplification arises from MCS interactions (e.g.~\citeA{Schwendike2010convection}). Both measures also point to rapid decay of the composite MCS--AEW system from the evening of 17/08/2016 onwards, which coincides with when the MCS cores begin to transition offshore. By the following morning (Fig.~\ref{fig:case_study}e), the MCS anvil has split up, decoupled from the AEW trough, and is associated with only weak upscale energy transfer at all $L$ and pressure levels (Fig.~\ref{fig:case_energetics}). Consequently, the wave trough CV is half of its peak value (bottom panel,~\ref{fig:case_study}e) and only weakly visible from streamlines. At 600\,hPa (Fig.~\ref{fig:case_study}f) there remains weak upscale energy transfer at $L$=1000\,km within the trough, which exists as a remnant from nocturnal MCS activity. Figure~\ref{fig:case_energetics} shows that the decline in EKE tendency again lags the 200\,hPa, $L$=500\,km $\iset$ signal, and occurs in phase with the $L$=1000\,km and $L$=2000\,km signals. We therefore conclude that the decay of the wave's embedded MCS, and thus the decreased upscale transfer of energy to the trough, contributes to the decay of the AEW.

This case study encapsulates the strong correlation between MCS--associated upscale energy transfer and AEW growth, but also highlights many other open questions that arise from studying $\iset$. In the mid--level, $L$=1000\,km fields for example (right column), we find significant bands of interscale energy transfer far from MCS activity. The dynamics ahead of AEW troughs are particularly intricate, with secondary moist mesoscale circulations, supported by MCS diabatic PV generation, extending westwards into dry, northerly flows \cite{Tomassini2017interaction,Tomassini2018mesoscale}. Examining spatial maps throughout the Explicit simulation suggests that the prominent 600\,hPa upscale transfer bands (e.g. Fig.~\ref{fig:case_study}d) always come at the leading edge of wave troughs, but also extend over 1000\,km beyond the trough and correlate with similarly--sized extents of anomalously dry air. This points towards a wider role of northerly mid--level Saharan air layer intrusions, which play a key role in the meteorology of the Sahel \cite{Parker2017meteorology} and can enhance MCS intensity and contribute to mid--level PV generation \cite{Tomassini2018mesoscale}. In contrast, the downscale region north and east of the trough in Fig.~\ref{fig:case_study}d corresponds to positive moisture anomalies advected by the southerly monsoon flow. 

It is also noteworthy that the wave trough's zonal propagation can be seen to slow down markedly during its peak EKE amplification phase; the case study wave trough's median zonal velocity of -5.61 m\,s$^{-1}$ across the onshore domain shown in Fig.~\ref{fig:case_study} is 1.16 m\,s$^{-1}$ slower than the equivalent for the remaining population complement. Strong PV dipoles generated by diabatic heating are predicted to marginally decrease the propagation of baroclinic waves \cite{de2010baroclinic}. The AEW's deceleration therefore provides a further indication that the upscale energy transfer contributing to the EKE tendency is caused by upwards diabatic transport of MCS--generated PV. Understanding this mechanism in detail is a crucial topic for future work that we will return to in our Conclusions. Here however, we return to the equivalent parameterised simulation. The equivalent Param strongest--wave case shows no significant correlation between 600\,hPa upscale energy transfer and wave EKE tendency (Fig.~S\SuppRef{}d). We do however find similar correlations between broad, non--AEW trough 600\,hPa upscale transfer regions and dry air intrusions in Param, but far weaker wave--centred signals during AEW growth. This motivates us returning to contrast our results from Explicit with dynamics in Param.


\section{Regional energetics with parameterised and explicit convection}
\label{sec:energetics}

We have so far focussed our discussion on processes within the Explicit simulation, which is known to show more realistic characteristics than Param for West African MCS dynamics \cite{Maybee2024wind} and convection--circulation couplings \cite{Bassford2026tropical}. Contrasting the energetics of AEWs in Explicit and Param, however, provides an enlightening insight into the active role played by upscale energy transfer from MCSs in the convection permitting simulation.

\begin{figure}[t]
    \centering
    \includegraphics[width=0.9\textwidth]{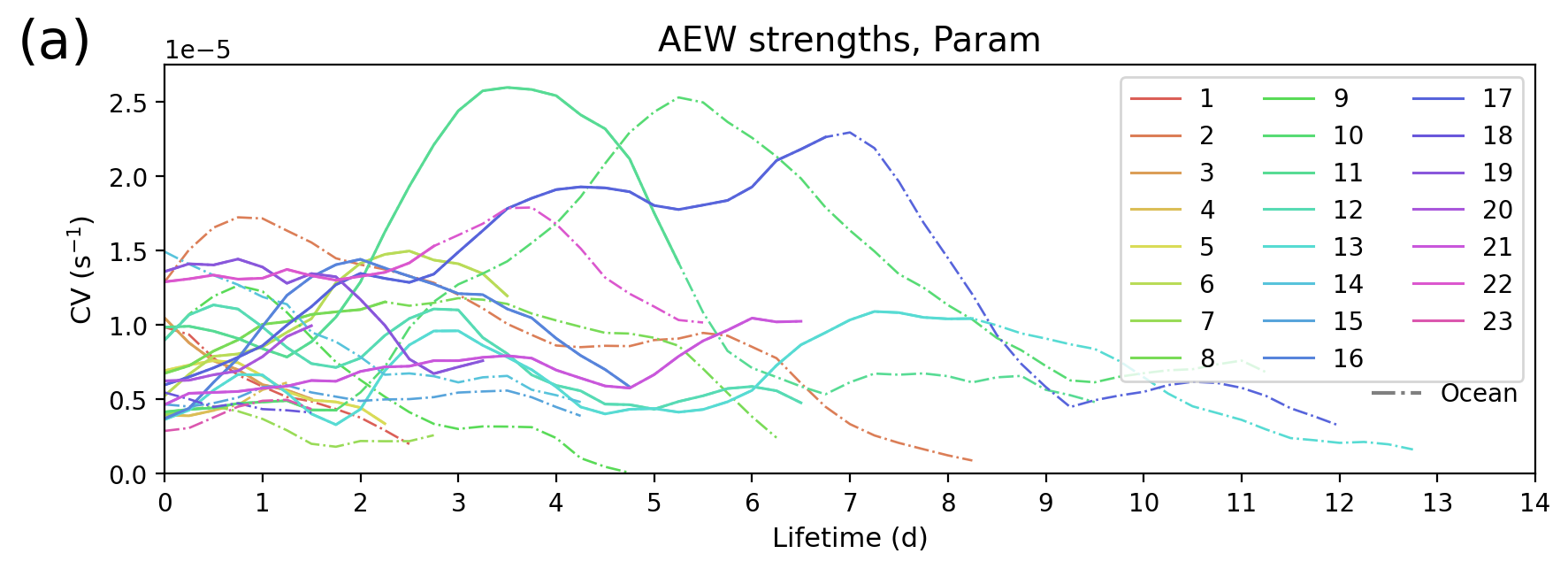}
    \includegraphics[width=\textwidth]{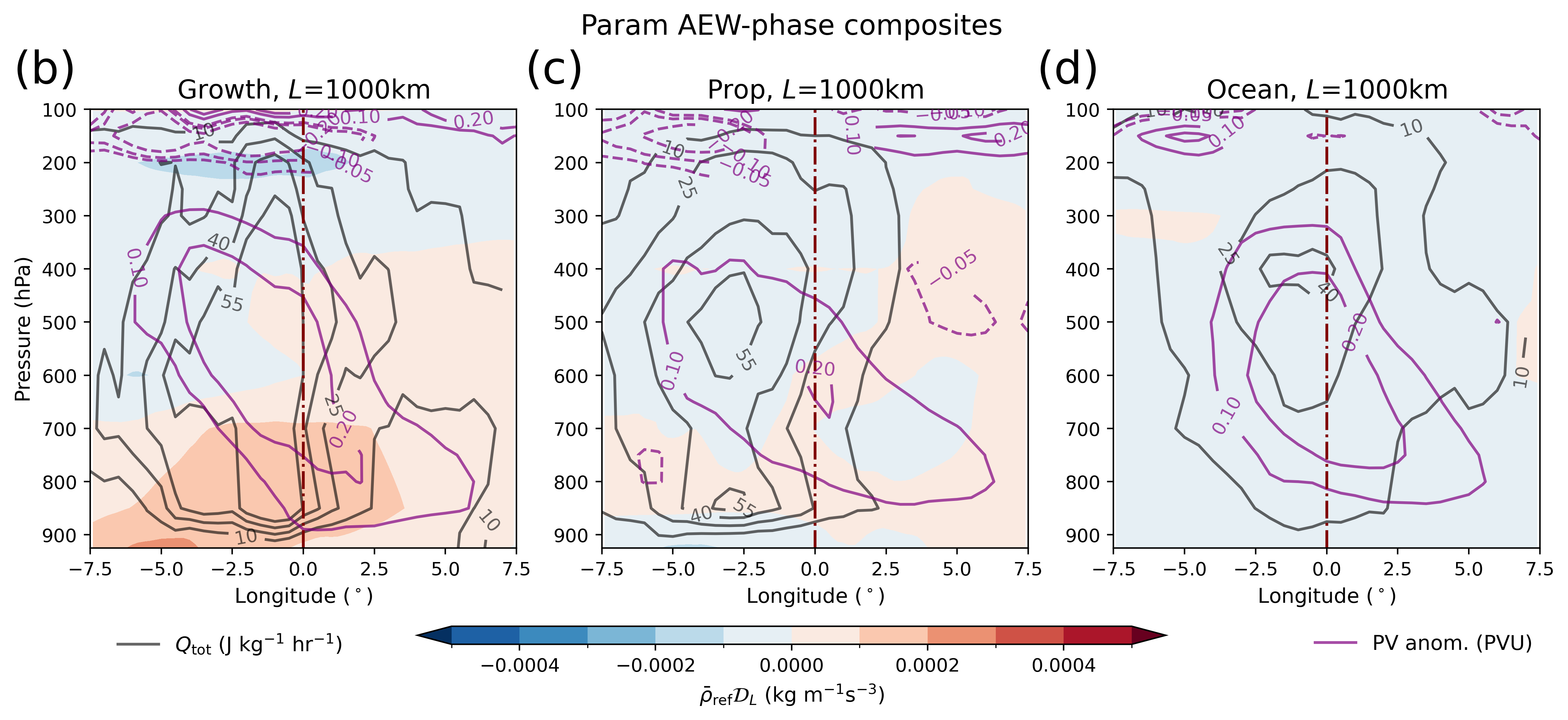}
    \caption{\textit{(a)} Grouped plots of individual evolution of AEW strength for each tracked system in Param post--wave initiation, plotted as for Explicit in Fig.~\ref{fig:hovmoellers}d. \textit{(b--d)} AEW track--centred composite vertical sections in Param of $L$=1000\,km mass--weighted $\iset$ (shading), diabatic heating (contours) and PV anomalies (purple) for growing, propagating and oceanic phase AEWs, respectively, plotted as for Explicit in Fig.~\ref{fig:aewISETcomps}.}
    \label{fig:param_lossett}
\end{figure}
Figure~\ref{fig:param_lossett}a shows the evolution of AEW strength for all tracked wave troughs in Param; Hovm\"{o}ller diagrams are given in Fig.~S7. Comparing the wave population to the results for Explicit (direct counterpart, Fig.~\ref{fig:hovmoellers}d), we immediately see that wave strengths are notably stronger overall in Param versus Explicit. Indeed, the median tracked trough CV is 43\% larger in Param (7.5$\times$10$^{-6}$ s$^{-1}$ versus 5.2$\times$10$^{-6}$ s$^{-1}$ in Explicit). The second--strongest trough in Param is oceanic (dotted lines), contrasting with the oceanic decays seen in the Explicit population. Of course the wave populations are small, being derived from only 40 day simulations. Nevertheless, the difference is consistent with the stronger AEJ in Param (Fig.~S1), which is itself consistent with results derived from longer global simulations using the same physics configurations \cite{Maybee2026dynamics}.

This then suggests an apparent contradiction: from analysis of Explicit in previous sections, we have emphasised the role of upscale energy transfer in AEW growth, but now find stronger wave troughs in a model with weaker mesoscale upscale energy transfers (Fig.~\ref{fig:ISETspectra} and~\ref{fig:regionalISETbands}). Indeed, local analysis of wave--trough centred composites confirms that there is a far weaker correlation between interscale energy transfer and AEW phase in Param than in Explicit. Figs.~\ref{fig:param_lossett}b--d show composite mass--weighted $\iset$ at $L$=1000\,km, as these results are representative of the wider scale range plotted for Explicit in Fig.~\ref{fig:aewISETcomps}. All values are notably weaker than for Explicit, and the only apparent signature of an MCS leading the wave trough is in the growth phase (Fig.~\ref{fig:param_lossett}c). We again find vertical PV dipoles collocated with diabatic heating that leads the AEW trough in all phases, but the heating profile appears less top--heavy for growing AEWs. The positive PV anomalies are more tilted than in Explicit, reflecting the vertical shear of the stronger AEJ in Param.

\begin{figure}[h!]
    \centering
    \includegraphics[width=\textwidth]{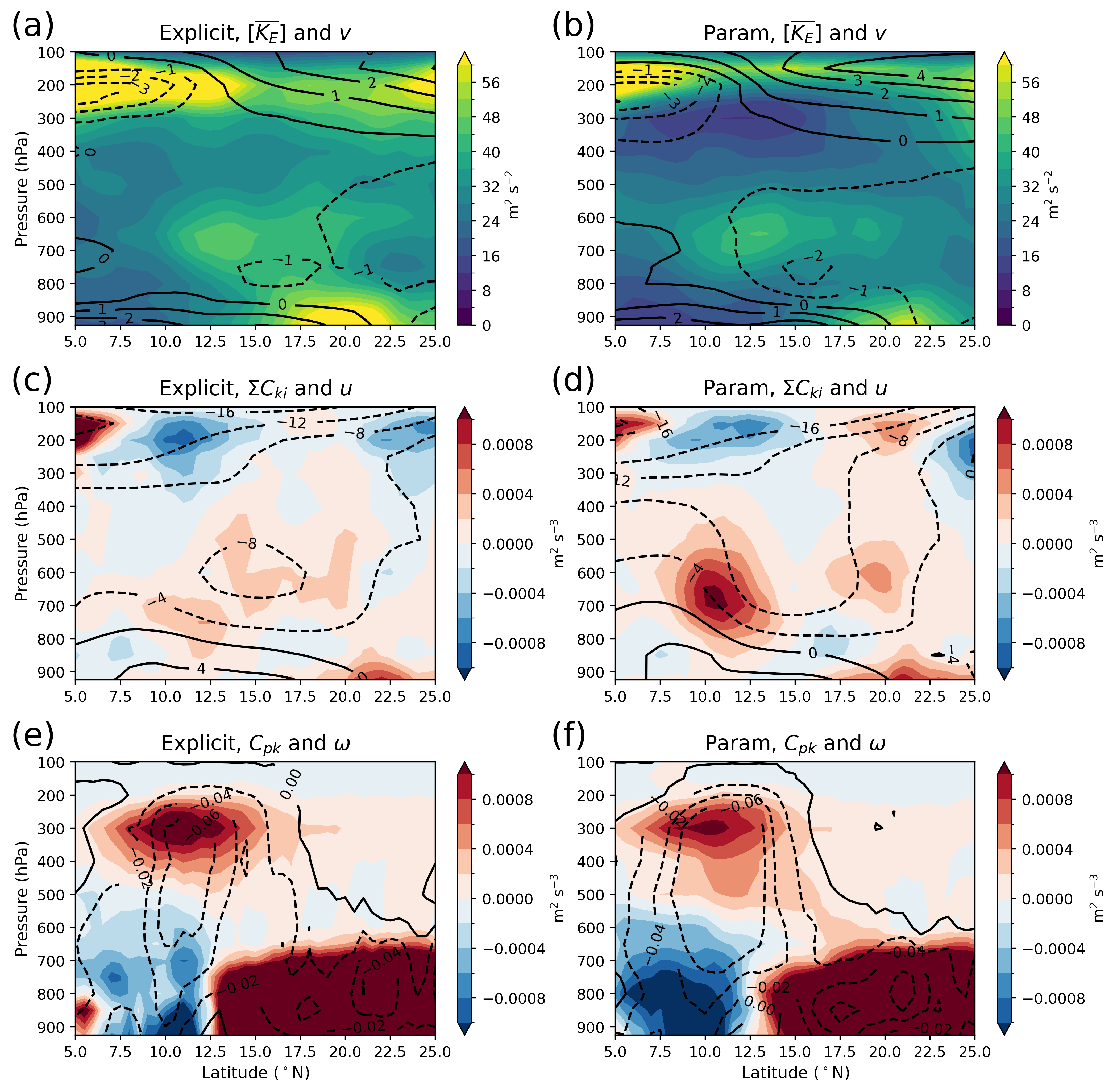}
    \caption{Full--period zonal mean selected energetics terms (-15$^\circ$--20$^\circ$E) in Explicit (left column) and Param (right column). Top row \textit{(a,b)} shows EKE (shading) and meridional wind (contours, units m\,s$^{-1}$; second row \textit{(c,d)} barotropic energy conversion $\Sigma_{i=1}^4C_{ki}$ (shading) and zonal wind (contours, units m\,s$^{-1}$); and bottom row \textit{(e,f)} baroclinic energy conversion $C_{pk}$ (shading) and vertical velocity (contours, units Pa\,s$^{-1}$). All plots are calculated from 2--8 day wavelet--filtered fields and the bottom two rows share common colourbar scales.}
    \label{fig:energetics}
\end{figure}

Despite their weaker composite $\iset$ signatures, AEWs remain stronger in Param. To explore the different processes driving wave growth in the two models, we step back from LoSSETT's calculation of interscale energy transfer and consider the zonal mean energetic contributions to AEW dynamics in the simulations. We follow standard convention by now taking perturbations in Eq.~\ref{eqn:eke} with respect to the zonal mean of the filtered winds. This is denoted by square brackets and taken over -15--20$^\circ$E to focus on signals over continental Africa; bars then indicate a temporal average over the full 40 days. Plotting the mean EKE for our simulations (Figs.~\ref{fig:energetics}a,b), we find local maxima in both simulations at 650\,hPa (around 13$^\circ$N) and 900\,hPa (21$^\circ$N); these are the southern and northern track AEW trains. Mean EKE is generally stronger in Explicit (panel a), especially at 200\,hPa, and the 600\,hPa southern--track maximum lies $\sim$2$^\circ$ further north on average. The strong 200\,hPa EKE extending to 15$^\circ$N in Explicit likely represents the simulation's enhanced MCS activity, with tight temporal coupling of their strong divergent outflows to AEW variability.

Following the conventions of \citeA{Hsieh2007study}, the governing equation for the zonal--mean EKE is
\begin{equation}
    \frac{\partial}{\partial t}[\overline{K_E}] = \sum_{i=1}^4C_{k\,i} + C_{pk} - D_E + K_{EB} + \Phi_{EB}\,.
\end{equation}
Here the terms on the RHS are barotropic energy conversion; baroclinic conversion; frictional dissipation; the boundary flux of $[\overline{K_E}]$; and boundary pressure work done by eddies. \citeA{Hsieh2007study} showed that barotropic and baroclinic energy conversion dominate the generation of (3--5 day eddy filtered) $K_E$ over West Africa, and moreover our interest is in the southern track AEW train in the centre of our domain. We therefore ignore the boundary terms in what follows. $D_E$ cannot be calculated from available output, but dissipative terms always act to reduce EKE and can be ignored for consideration of positive contributions to AEW energetics.

The summation term encodes the energy conversion from shear of the horizontal flow components. In typical notation, these are considered separately and take the form
\begin{align}
    C_{k1}+C_{k2}&=-\overline{[u'v']\frac{\partial[u]}{\partial y}} - \overline{[u'\omega']\frac{\partial[u]}{\partial p}}\,,\\
    C_{k3}+C_{k4}&=-\overline{[{v'}^2]\frac{\partial[v]}{\partial y}} - \overline{[v'\omega']\frac{\partial[v]}{\partial p}}\,,
    \label{eqn:barotropic}
\end{align}
respectively. We plot both combined since the zonal winds are much stronger over our domain. The AEJ is stronger in Param than Explicit (Fig.~S1), and we would therefore expect stronger barotropic instabilities associated with the zonal wind. Figs.~\ref{fig:energetics}c,d show that barotropic energy generation indeed drives much stronger EKE production in Param than Explicit. The contribution of the zonal wind ($C_{k1}+C_{k2}$) dominates the meridional terms ($C_{k3}+C_{k4}$; not shown). Thus in Param, barotropic energy generation peaks at 650\,hPa just south of the zonal mean jet (black contours). As shown by the zonal mean EKE (Fig.~\ref{fig:energetics}b), this maximum in energy generation aligns with the location of the southern track AEW train that our paper focuses on. There is only a weak equivalent local maximum in Explicit. Of the contributing terms to $\Sigma C_{ki}$, barotropic energy production by horizontal shear dominates ($C_{k1}$) the mid--levels in both simulations (not shown). Near the AEJ the signs of the two zonal terms superpose, whereas at the tropopause they are opposite (and serve to cause an overall near--cancellation); the 10$^\circ$, 200\,hPa maximum in Explicit (Fig.~\ref{fig:energetics}c) is caused by enhanced upper--level horizontal shear, likely stemming from the enhanced MCS outflows aloft.

The other significant driver of EKE production is baroclinic energy conversion,
\begin{equation}
    C_{pk}=-\frac{R_d}{p}\overline{[\omega'\,T']}\,,
\end{equation}
i.e. the conversion of potential to kinetic energy. This is typically identified as the dominant term in the production of regional EKE, and indeed we find larger values than for barotropic energy conversion (Figs.~\ref{fig:energetics}e,f). However, there are few pronounced differences between the two simulations around the 650\,hPa location of the southern track AEWs in which we are primarily interested. Baroclinic energy production at the tropopause is stronger in Explicit than Param, and near the surface the significant positive region shown by both simulations extends further south into the Sahel. Both of these features can be interpreted through the lens of MCS differences. The mean ascent region over West Africa (negative $\omega$ contours, black) is more confined, further north and stronger at upper levels in Explicit. Combined with enhanced temperature anomalies from convective heating aloft, and boundary layer cooling via cold pool generation, the mean ascent pattern leads to enhanced creation of EKE in Explicit. The strong positive $C_{pk}$ in Explicit near the tropopause contributes significantly to the strong upper--level EKE values (Fig.~\ref{fig:energetics}a).

We can thus understand the stronger AEW troughs in Param versus Explicit as emerging from increased zonal barotropic energy production $C_{k1}+C_{k2}$ driving the AEW train as a consequence of the stronger simulated AEJ. The AEJ jet core extends further offshore in Param (Fig.~S1), providing a further explanation for the growing and strengthened oceanic AEWs relative to Explicit.

The interscale kinetic energy transfer computed by LoSSETT is, in many senses, a calculation (see Eq.~\ref{eqn:iset}) of energy production from shear terms across spatial scales \cite{YaoEtAl2024}. At the appropriate $L$, the zonal mean barotropic EKE production from the Lorenz energy cycle should be approximately equivalent to the zonal mean $\iset$. Our domain of interest is in essence a narrow channel, with a width of 1500\,km. Comparing Fig.~\ref{fig:energetics}c,d to the $\iset$ zonal means in Figs.~\ref{fig:ISETspectra} (Explicit) and Fig.~S4 (Param), there are indeed strong similarities between zonal mean $\Sigma C_{ki}$ and $\iset$ at $L$=1000\,km, with upper--level upscale energy transfer correlating with negative barotropic energy tendency, and mid/low--level downscale transfer correlating with barotropic energy production. In particular, the strong mid--level barotropic energy production by the AEJ in Param collocates with strong downscale transfer in the zonal mean $\iset$ in Fig.~S4b. 

The strong upper--level $L$=1000\,km upscale energy transfer in the Explicit zonal mean (Fig.~\ref{fig:ISETspectra}b) instead correlates with increased baroclinic energy production in the Lorenz cycle analysis versus Param. Note that strong baroclinic energy production in the boundary layer correlates to downscale transfer, showing the limited relationship between these terms. Nevertheless, the zonal mean energetics provide a further insight to our discussion of the regional energy transfer spectra in Sec.~\ref{sec:aew_length_scales}. The qualitative comparison of the Explicit spectrum (Fig.~\ref{fig:ISETspectra}d) to two--layer geostrophic turbulence models relied on a baroclinic energy conversion at upper--levels and barotropic mid--level conversion. Figure~\ref{fig:energetics} shows that these conditions are met in both simulations, and moreover that the baroclinic production of EKE at AEW latitudes correlates with where MCSs input upscale energy transfer into the zonal mean, and that barotropic energy conversion correlates with the AEW trains. Two--layer geostrophic turbulence provides a conceptual model for understanding the transfer of energy between these modes, but with only limited applicability to the AEW spectrum, as emphasised in Sec.~\ref{sec:aew_length_scales}. It would be valuable to explore such models using new theoretical descriptions of AEWs that capture their moist dynamics (e.g.~\citeA{Adames2026rotation}). Our results here suffice to demonstrate that incorporating the $\iset$ metric into AEW energetics enables deeper understanding of physical sources of zonal mean energetics terms, with the advantage of undertaking spatially localised analysis, going beyond zonal means.


\section{Discussion and Conclusions}
\label{sec:conclusions}

In this paper, we have sought to provide a novel perspective on the multiscale interactions between MCSs and African easterly waves by quantifying local interscale transfers of atmospheric kinetic energy over West Africa and the East Atlantic. Spatially localised interscale energy transfer $\iset$ has been calculated across a range of length scales $L$ using LoSSETT \cite{Mckinnon2025representation}, while for atmospheric profile data we have used two 40 day, $\sim$5km grid Met Office Unified Model simulations of Earth's tropics. The convection permitting configuration, Explicit, provides a more realistic representation of West African climate \cite{Maybee2024wind} than a counterpart with parameterised convection (Param), and shows stronger upscale energy transfer across scales, but particularly at scales associated with MCSs.

Our focus has been on using local quantification of interscale energy transfer to diagnose the upscale role of MCSs in West African climate, and their role in driving AEW growth. We have therefore focused our analysis on the Explicit simulation, using Param to provide complementary insights into physical processes rather than for model evaluation.  In Explicit, energy transfer composited about MCSs exhibits a strong characteristic dipole pattern for values of $L$ at MCS scales, with boundary layer downscale transfer and stronger upscale energy transfer at upper levels. This upscale energy transfer corresponds to the net divergent outflow from deep convection, is bounded by MCS anvils, and extends to transfer length scales greater than 1000\,km (Figs.~\ref{fig:mcsISETcomps} and~\ref{fig:ISETspectra}). Upscale energy generation by MCSs also behaves in a manner consistent with other measures of MCS intensity and in particular with MCS diabatic heating, increasing in favourable moist and strongly sheared convective environments (Fig.~\ref{fig:mcsISETcomps}c). Analysis of regional energy spectra shows convection contributes to upscale energy transfer from convective to synoptic scales at 200\,hPa, and that the peak upscale energy transfer coincides with the transition from downscale to upscale transfer at the 600\,hPa AEW level (Fig.~\ref{fig:ISETspectra}d), which itself occurs around the AEW Rhines and Rossby length scales.

We objectively tracked AEWs in the MetUM simulations from 700\,hPa curvature vorticity at 1$^\circ$ resolution using the QTrack algorithm \cite{Lawton2022influence}, using vorticity tendency and location criteria to divide AEW lifecycles into growth, propagation and ocean phases. Analysis of Hovm\"{o}ller diagrams and wave trough centred composites confirms that the Explicit simulation generates realistic AEW structures and propagation characteristics, but with wave strengths weaker than those in Param. We find that AEW troughs in Explicit are preceded by upper--level upscale energy transfer at scales up to 2000\,km, and by mid--level energy transfer that is downscale in the MCS scale range, but upscale for scales around those of observed AEW wavelengths. Vertical composites of $\iset$ about the wave phases (Fig.~\ref{fig:aewISETcomps}) reveal upscale energy transfer follows the characteristic MCS dipole pattern and is strongest for growing AEWs, especially at the  trough scale of $\sim$1000\,km. Case study analysis of the strongest simulated AEW in Explicit shows upscale energy transfer from MCSs can contribute to an increase in AEW eddy kinetic energy, and can occur at the wave--trough level for large $L$ (Fig.~\ref{fig:case_energetics}). Persistent upscale forcing from a large MCS embedded at the trough's leading edge acts as the source and contributes to exceptional growth (Fig.~\ref{fig:case_study}), while the case AEW also rapidly decays as soon as this MCS detaches and dissipates.

Our results support recent perspectives concluding that diabatic processes within MCSs are critical to the growth and maintenance of AEWs. \citeA{Russell2020african} argued that mid--level PV generation by MCS stratiform decks are the primary source of positive PV tendency for growing AEWs (see also \citeA{Tomassini2017interaction,Russell2020potential}). Our composite analysis of growing AEWs (Fig.~\ref{fig:aewISETcomps}a--c) shows that up to $L$=1000\,km, upscale energy transfer extends across MCS stratiform regions, correlates with diabatic heating profiles, and coincides with the negative--anomaly component of a vertical PV dipole. The absolute PV is always positive, and thus the composites are consistent with upscale transfer stemming from PV dilution due to diabatic mass transport \cite{Harvey2020diabatic,Saffin2021circulation}. Further investigation is essential for determining the precise mechanisms underpinning the upscale signatures we find leading AEWs, but our results begin to answer the question raised by \citeA{Russell2020african} of how heating from $L\sim$250\,km MCSs scales up to influence $L>$1000\,km AEW troughs: upscale energy transfer provides a physical pathway for this influence. 

Our study of simulations with explicit and parameterised convection also acts as a natural experiment into different processes contributing to AEW growth. In Param, we find little role for upscale energy transfer in AEW growth (Fig.~\ref{fig:param_lossett}), with traditional zonal mean energetics analysis showing barotropic generation of EKE is instead the primary driver. This term is much larger than in Explicit due to the stronger mid--level African Easterly Jet in Param, which drives an overall intensification of the AEW population. Our analysis emphasises the complex interactions which drive and maintain AEWs and does not negate the key role of baroclinic and barotropic instability in AEW dynamics (e.g. Fig.~\ref{fig:aew_comps}d). We have shown that upscale energy transfer from MCS heating also contributes to AEW growth, and that its simulation is critically dependent on the representation of deep convection. An ambitious promise of global km--scale convection permitting models such as Explicit is to explicitly simulate the upscale impacts of convection on circulations, providing benefit to the representation of the largest scales as well as local extremes. Several studies already suggest this can be achieved \cite{Marsham2013role,Pante2019resolving,Rixen2026grid}. Our study demonstrates process--analysis methods that can be deployed to support this goal, and moreover shows the potential benefit of this modelling programme for advancing theoretical understanding of our atmosphere.

Finally, it is important to note that this paper forms one of several beginning to apply the methods of interscale energy transfer, long established in turbulence theory, to local phenomena represented by full--complexity models of Earth's atmosphere. There remains much to explore with this metric \cite{Faranda2018computation,Kouhen2024convective,kouhen2024,Lojko2026significant,Sichani2026atmospheric}. For example, understanding how the diabatic parameters explored in this study apply to other tropical phenomena which show upscale energy transfer, such as equatorial waves \cite{Mckinnon2025representation}. Moreover, the scale--sensitive energetic connection between equatorial waves and AEWs would also serve as a logical extension to our paper, connecting back to the work of \citeA{Yang2018linking}, who showed amplifying AEWs have a strong relationship with equatorial westward mixed Rossby--gravity waves. In this paper, we have only been able to correlate upscale energy transfer, vertical heating profiles and diabatic PV structures. At present we lack the theoretical basis to relate interscale energy transfer, the language of turbulence theory, with potential vorticity tendencies, the language of dynamical meteorology. A complete theory for turbulence and interscale energy transfer in tropical modes such as AEWs does not yet exist. \citeA{Prince2022convectively} have shown deformation--driven thinning of negative PV during Hurricane Irma caused upscale energy transfer, but beyond this case study we lack a unified, mechanistic understanding of interscale energy transfers. We hope our paper motivates future research into this field.


\section*{Acknowledgements}

BM acknowledges funding from the NERC Hurac\'{a}n grant NE/W009587/1, while  EMG acknowledges support from SCENARIO DTP NERC Grant NE/5007261/1 and Met Office CASE funding. DS and JM are supported by the UPSCALE project, funded by the UK Department for Science, Innovation \& Technology (DSIT) National Capability AI Programme and managed by the Met Office. The authors thank Leo Kelly for sharing results derived during his MRes studies; Richard W. Jones for reviewing details of the Met Office simulations; and Conni Klein, John Marsham, Fran Morris and Juliane Schwendike for related conversations and comments.

\section*{Data Availability Statement}
Code and data underpinning this research are available publicly at \url{https://github.com/BMaybee/AEW\_upscale\_transfer} and  \url{https://github.com/ElliotMG/LoSSETT}.

\section*{Supporting Information}

Supporting Information can be found with the online version of this article.

\bibliography{references}

\end{document}